\documentclass[journal]{IEEEtran}
\ifCLASSINFOpdf
\else
   \usepackage{graphicx}
   \graphicspath{{../eps/}}
   \DeclareGraphicsExtensions{.eps}
\fi

\usepackage{amsmath}
\usepackage{amsfonts,amssymb}
\usepackage{amssymb}
\usepackage{wrapfig}
\usepackage{psfrag}
\usepackage{epstopdf}
\usepackage{cite}
\usepackage{graphicx}
\usepackage{subfigure}
\usepackage{threeparttable}
\usepackage{cases}
\usepackage{subeqnarray}
\usepackage{color}
\usepackage{underscore}
\usepackage{verbatim}
\usepackage{bm}
\usepackage{stfloats}
\usepackage{xpatch}

\newtheorem{lemma}{Lemma}
\newtheorem{algorithm}{Algorithm}
\newtheorem{proposition}{Proposition}

\usepackage{algorithm}
\usepackage{algorithmic}

\begin{document}
\title{Wireless Communication with Flexible Reflector: Joint Placement and Rotation Optimization for Coverage Enhancement}
%
%
%%
%%
%% author names and IEEE memberships
%% note positions of commas and nonbreaking spaces ( ~ ) LaTeX will not break
%% a structure at a ~ so this keeps an author's name from being broken across
%% two lines.
%% use \thanks{} to gain access to the first footnote area
%% a separate \thanks must be used for each paragraph as LaTeX2e's \thanks
%% was not built to handle multiple paragraphs
%%
%
\author{
        Haiquan~Lu,
        Zhi~Yu,
        Yong~Zeng,~\IEEEmembership{Fellow,~IEEE,}
        Shaodan~Ma,~\IEEEmembership{Senior Member,~IEEE,}\\
        Shi~Jin,~\IEEEmembership{Fellow,~IEEE,}
        and
        Rui~Zhang,~\IEEEmembership{Fellow,~IEEE}
        % <-this % stops a space
\thanks{This work was supported in part by the Natural Science Foundation for Distinguished Young Scholars of Jiangsu Province under Grant BK20240070, in part by the National Natural Science Foundation of China under Grant 62071114, and in part by the Fundamental Research Funds for the Central Universities under Grant 2242022k60004. (\emph{Corresponding author: Yong Zeng.}) }
\thanks{Haiquan Lu, Zhi Yu, Yong Zeng, and Shi Jin are with the National Mobile Communications Research Laboratory and Frontiers Science Center for Mobile Information Communication and Security, Southeast University, Nanjing 210096, China. Haiquan Lu and Yong Zeng are also with the Purple Mountain Laboratories, Nanjing 211111, China (e-mail: haiq_lu@163.com, \{zhiyu, yong_zeng, jinshi\}@seu.edu.cn). }
\thanks{Shaodan Ma is with the State Key Laboratory of Internet of Things for Smart City and the Department of Electrical and Computer Engineering, University of Macau, Macao SAR, China (e-mail: shaodanma@um.edu.mo).}
\thanks{Rui Zhang is with School of Science and Engineering, Shenzhen Research Institute of Big Data, The Chinese University of Hong Kong, Shenzhen, Guangdong 518172, China (e-mail: rzhang@cuhk.edu.cn). He is also with the Department of Electrical and Computer Engineering, National University of Singapore, Singapore 117583 (e-mail: elezhang@nus.edu.sg).}

% <-this % stops a space
}

\maketitle

\begin{abstract}
 Passive metal reflectors for communication enhancement have appealing advantages such as ultra low cost, zero energy expenditure, maintenance-free operation, long life span, and full compatibility with legacy wireless systems. To unleash the full potential of passive reflectors for wireless communications, this paper proposes a new passive reflector architecture, termed \emph{flexible reflector} (FR), for enabling the flexible adjustment of beamforming direction via the FR placement and rotation optimization. We consider the multi-FR aided area coverage enhancement and aim to maximize the minimum expected receive power over all locations within the target coverage area, by jointly optimizing the placement positions and rotation angles of multiple FRs. To gain useful insights, the special case of movable reflector (MR) with fixed rotation is first studied to maximize the expected receive power at a target location, where the optimal single-MR placement positions for electrically large and small reflectors are derived in closed-form, respectively. It is shown that the reflector should be placed at the specular reflection point for electrically large reflector. While for area coverage enhancement, the optimal placement is obtained for the  single-MR case and a sequential placement algorithm is proposed for the multi-MR case. Moreover, for the general case of FR, joint placement and rotation design is considered for the single-/multi-FR aided coverage enhancement, respectively. Numerical results are presented which demonstrate significant performance gains of FRs over various benchmark schemes under different practical setups in terms of receive power enhancement.
\end{abstract}


\IEEEpeerreviewmaketitle
% >>>>>>>>>>>>>SECTIONS I -  here >>>>>>>>>>>>
\section{Introduction}
 The evolution of mobile communication networks has witnessed the tremendous success of multi-antenna technology, such as multiple-input multiple-out (MIMO) widely used in the fourth-generation (4G) wireless networks and massive MIMO in the fifth-generation (5G) wireless networks, due to their substantial spatial diversity and multiplexing gains. For the forthcoming sixth-generation (6G) era, massive MIMO is expected to evolve towards extremely large-scale MIMO (XL-MIMO) via scaling up the antenna number by an order of magnitude \cite{lu2024tutorial,bjornson2019massive,wang2024extremely}, e.g., hundreds or even thousands of antennas at the base station (BS), so as to support the ambitious capabilities required by 6G \cite{ITU}. Despite achieving the unprecedented improvement in the spatial resolution and spectral efficiency, XL-MIMO also faces practical challenges such as expensive hardware cost and high energy expenditure \cite{lu2024tutorial,Han2023Towards}. To address such issues, there has been an upsurge of interest in exploiting various sparse array architectures, including uniform sparse array \cite{wang2023can,wang2024enhancing,lu2024group} and non-uniform sparse array, such as modular, nested, and co-prime arrays \cite{li2024sparse,pal2010nested,wang2016coarrays}. Compared to the conventional compact array with neighboring elements separated by half wavelength, sparse arrays can achieve a larger array aperture by configuring the antenna spacing  larger than half wavelength, without increasing the number of antenna elements. This thus provides improved spatial resolution and degree-of-freedom (DoF) for enhancing both communication and sensing performances.

 Besides sparse array, there was extensive research on various cost-effective and energy-efficient hardware architectures for MIMO, such as analog beamforming, lens antenna array, and low-resolution analog-to-digital converter (ADC) \cite{zeng2024tutorial}. More recently, another promising approach, namely intelligent reflecting surface (IRS) or reconfigurable intelligent surface (RIS) aided communication has been proposed \cite{wu2021intelligent,di2020smart,lu2021aerial,tang2020wireless}. Specifically, without any active radio-frequency (RF) chains, IRS is a metasurface consisting of a large number of passive reflecting elements, which are capable of configuring the wireless propagation environment proactively via adjusting their amplitude and/or phase shifts. However, the cost of IRS/RIS increases significantly with its size or number of reflecting elements, especially for that operating at higher frequency bands, which is still practically formidable for large-scale deployment of IRS/RIS in wireless networks.

 To further reduce the hardware cost and energy expenditure, fully passive metal reflectors, usually made of copper, aluminum or conductive coating, were also introduced and have gained increasing interest. Initially, the passive reflector was proposed as an alternative of active satellite relays, due to its simpler operation and maintenance \cite{ryerson1960passive,cutler1965passive}. In the field of antenna design, different types of passive reflectors, such as plane, corner and parabolic reflectors, were utilized to fabricate the reflector antenna, which has been widely applied to the radio astronomy, satellite tracking, and remote sensing \cite{rahmat2015reflector,balanis2016antenna}. Recently, passive metal reflectors were introduced in wireless communications, aiming at signal/coverage enhancement, channel rank/diversity improvement, interference rejection, secure transmission, and so on. For example, to compensate for the severe penetration loss suffered by millimeter wave (mmWave) communications, passive metal reflectors were deployed to enhance the signal coverage in both indoor \cite{huang2004investigation,chan20153d,han2017enhancing,khawaja2020coverage} and outdoor scenarios \cite{khawaja2020coverage,hager2023holistic}. Moreover, from the perspective of network planning, the deployment of passive reflectors helps to reduce the number of BSs required for area coverage, by jointly designing the placement positions of BSs and passive reflectors \cite{anjinappa2021base}. This thus significantly reduces the deployment cost and energy expenditure, as compared to conventional approaches such as increasing transmit power, the number of BSs, access points (APs), relays, and/or antennas \cite{khawaja2020coverage}. Another appealing benefit for deploying passive reflectors is the MIMO channel rank/diversity improvement, thanks to the creation of additional strong multi-path components \cite{anjinappa2021base,singh2023stabilizing,yu2023wireless}. On the other hand, instead of facilitating the connectivity between desired nodes, passive reflectors can also be deployed to physically block the interference from undesired nodes \cite{barreiro2006passive}. Besides, customized wireless communication environment for enhanced security and privacy can be achieved by placing the reflector to block the channel from the legitimate transmitter to malicious users \cite{chan20153d,han2017enhancing}.

 Note that compared to active BSs/APs or semi-passive IRSs, fully passive metal reflectors possess promising advantages such as ultra low cost, zero energy expenditure, maintenance-free operation, long life span, and full compatibility with legacy wireless systems \cite{anjinappa2021base,yu2023wireless}. Moreover, passive reflector is of high compatibility and efficient scalability, which can be integrated into existing and future wireless networks transparently, without changing the network protocol. It is also worth mentioning that the reflection wave of the metal reflector exhibits a beam shape \cite{balanis2016antenna,ozdogan2020intelligent}, and many efforts have been devoted to mathematically modeling the signal reflection by passive metal reflectors. Specifically, by fixing its normal vector, the scattered field expression of a metal surface was derived in \cite{ozdogan2020intelligent}. Furthermore, by taking into account the factors of metal reflector size, orientation and polarization, a more general reflection model was derived in closed-form, followed by experiment measurements to verify its accuracy \cite{yu2023wireless}. Nevertheless, passive reflector does not have adjustable electronic components, and is thus unable to apply conventional active/passive beamforming to adjust the reflection signal amplitude and/or direction.

 The aforementioned works mainly deploy the passive metal reflectors with fixed placement and orientation. This implies that once the metal reflector is deployed, the signal scattered field is determined. Thus, anomalous reflection to any desired direction cannot be achieved in general, unlike the case of IRS. To fully unleash the potential of passive reflector, in this paper, we propose a new architecture for it, termed \emph{flexible reflector} (FR), where the placement position and rotation angle of metal reflector can be flexibly adjusted to dynamically alter its beamforming direction. In contrast to multi-antenna beamforming which is achieved by dynamically controlling the signal phase of each antenna element, or IRS reflective beamforming by controlling the phase shift of each reflecting element, FR adjusts the placement position and/or rotation angle to manipulate the specular direction, i.e., the beamforming direction. This thus enables a new paradigm of passive beamforming, without relying on RF chains, phase shifters or reflecting elements, nor complex signal processing. Moreover, the practical implementation of FR resembles existing movable antennas (MAs) with position adjustment \cite{zhu2024movable,zhu2024modeling,ma2024mimo,zhu2025tutorial} as well as six-dimensional movable antennas (6DMAs) with both position and rotation adjustment \cite{shao20246d,shao20246ddiscrete,shao20246dma,shao2025distributed}. Specifically, by mounting the reflector on a motor-driven rotatable shaft, and integrating the entire component onto a sliding platform, the rotation angle and placement position can be flexibly adjusted via the mechanical control \cite{ning2024movable,zheng2025rotatable}. In this paper, we consider a general multi-FR aided wireless communication system, where one-dimensional (1D) movement and rotation of FR are considered to maximize the minimum expected received signal power across all locations in a given target area. The main contributions of this paper are summarized as follows:

 \begin{itemize}[\IEEEsetlabelwidth{12)}]
 \item Firstly, we introduce a multi-FR aided wireless communication system, where a flexile adjustment of passive reflective beamforming direction is enabled by optimizing the placement position and rotation angle of each FR. By deriving the received signal power at any location in terms of each FR's placement and rotation angle, we formulate an optimization problem to maximize the minimum expected receive power across all locations in a given area, subject to the practical placement constraint to avoid overlap and signal blockage, as well as the rotation constraint to ensure the effective reflection.
 \item Secondly, to gain useful insights, we consider the special case of  movable reflector (MR) without rotation. For the special scenario of single-MR aided single target location power enhancement, the optimal MR placement position is derived in closed-form for electrically large or small reflector, respectively. It is analytically shown that the MR should be placed at the specular reflection point for electrically large reflector, and passive beamforming direction adjustment is enabled via placement position optimization. Then, for the scenario of multi-MR aided single target location power enhancement, the placement positions of multiple MRs are jointly designed such that the receiver is located within their beamforming main lobes. Moreover, for area coverage enhancement with a single-MR, the minimum array factor across all locations is derived in closed-form, and its optimal placement is obtained accordingly. Subsequently, a sequential placement algorithm is proposed for the multi-MR case.
 \item Thirdly, we consider the general case of FR with both placement position and rotation angle adjustments. Starting from the single-FR aided single target location power enhancement, we derive the optimal rotation angle for the FR with its placement position fixed in closed-form. Then, for the case of multi-FR, the minimum distance between any two FRs to avoid overlap and signal blockage is derived. Next, the joint placement and rotation design is proposed for the single-/multi-FR aided area coverage enhancement scenarios. Finally, extensive numerical results are presented to demonstrate the significant performance gains of FRs over various benchmark schemes under different practical setups.
 \end{itemize}

 The rest of this paper is organized as follows. Section~\ref{sectionSystemModel} introduces the system model and formulates the optimization problem to maximize the minimum expected receive power across all locations in a given target area. Section~\ref{sectionMR} studies the special case of the single/multi-MR aided coverage enhancement. In Section~\ref{sectionFR}, joint placement and rotation design is considered for the single-/multi-FR aided coverage enhancement by extending MR to FR. Section~\ref{sectionNumericalResults} provides the numerical results and relevant discussions, and this paper is concluded in Section~\ref{sectionConclusion}.

 \emph{Notations:} Scalars are denoted by italic letters. Vectors and matrices are denoted by bold-face lower- and upper-case letters, respectively. ${{\mathbb{C}}^{M \times N}}$ represents the space of $M \times N$ complex-valued matrices. For a vector ${\bf{x}}$, $\left\| {\bf{x}} \right\|$ denotes its Euclidean norm, and ${\bf{x}}^T$ denotes its transpose. The symbol ${\rm j}$ denotes the imaginary unit of complex numbers, with ${{\rm j}^2} =  - 1$. For real number $x$, $\left\lfloor x \right\rfloor $ and $\left\lceil x \right\rceil$ denote the floor and ceiling operations, respectively. ${\mathbb E}\left({\cdot}\right)$ denotes the statistical expectation.

% >>>>>>>>>>>>>SECTIONS II -  here >>>>>>>>>>>>
\section{System Model And Problem Formulation}\label{sectionSystemModel}
 \begin{figure}[!t]
 \centering
 \centerline{\includegraphics[width=3.2in,height=2.28in]{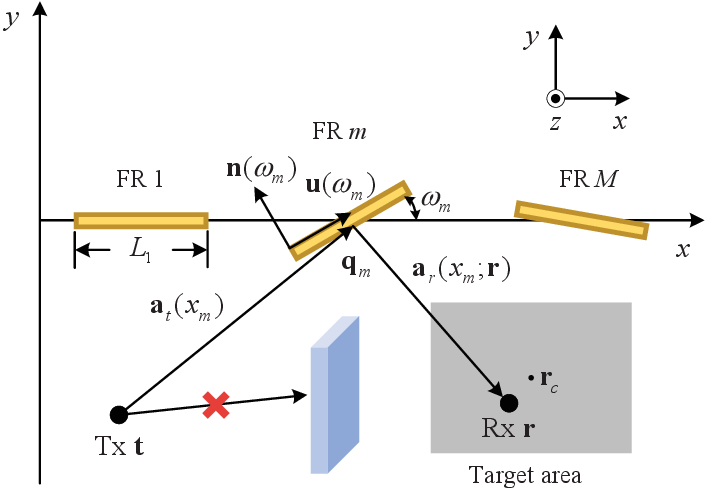}}
 \caption{Wireless communication enhanced by multiple FRs.}
 \label{fig:systemModel}
 \end{figure}

 As shown in Fig.~\ref{fig:systemModel}, we consider a multi-metal plate reflector aided wireless communication system, where $M$ FRs are deployed to assist in the communication from the transmitter (Tx) to the receiver (Rx) at arbitrary location in a given target area $\cal A$. The direct link from the Tx to the target area is assumed to be negligible due to blockage. Each reflector is assumed to have size ${L_1} \times {L_2}$, and is centered at the $x$-axis. The location of the Tx is ${\bf{t}} = {\left[ {{x_t},{y_t},0} \right]^T}$. The target area $\cal A$ is assumed to be a rectangular area on the $x$-$y$ plane, and its center is denoted as ${{\bf{r}}_c} = {\left[ {{x_c},{y_c},0} \right]^T}$. The length and width of $\cal A$ are $D_x$ and $D_y$, respectively. Thus, any location within $ {\cal A}$ can be denoted as ${{\bf{r}}} = {\left[ {{x_r},{y_r},0} \right]^T}$, with ${x_r} \in \left[ {{x_c} - {D_x}/2,{x_c} + {D_x}/2} \right]$ and ${y_r} \in \left[ {{y_c} - {D_y}/2,{y_c} + {D_y}/2} \right]$. Moreover, the center location of reflector $m$ is denoted as ${{\bf{q}}_m} = {\left[ {{x_m},0,0} \right]^T}$, $\forall m \in {\cal M}$, with ${\cal M} \triangleq \left\{1, \cdots, M\right\}$. Since the reflectors may move along the $x$-axis, the notation $x_m$ is used to denote the center location of reflector $m$. The distance between the Tx and the center of reflector $m$ is ${d_t}\left( {x_m} \right) = \left\| {{{\bf{q}}_m} - {\bf{t}}} \right\|$, and that between the center of reflector $m$ and any location $\bf r$ in ${\cal A }$ is ${d_r}\left( {{x}_m;{\bf{r}}} \right) = \left\| {{\bf{r}}- {\bf{q}}_m } \right\|$.

 Let ${\omega _m}$ denote the 1D rotation angle of reflector $m$, and the rotation matrix is given by \cite{diebel2006representing,shao20246d}
 \begin{equation}\label{rotationMatrix}
 {\bf{R}}\left( {{\omega _m}} \right) = \left[ {\begin{array}{*{20}{c}}
 {\cos {\omega _m}}&{ - \sin {\omega _m}}&0\\
 {\sin {\omega _m}}&{\cos {\omega _m}}&0\\
 0&0&1
 \end{array}} \right].
 \end{equation}

 Denote by ${\bf{u}}\left( {{\omega _m}} \right)$ the normalized direction vector along the edge of reflector $m$ under given rotation angle ${\omega _m}$, and ${\bf{n}}\left( {{\omega _m}} \right)$ the corresponding normal direction vector, as illustrated in Fig.~\ref{fig:systemModel}. Thus, we have
 \begin{equation}
 {\bf{u}}\left( {{\omega _m}} \right) = {\bf{R}}\left( {{\omega _m}} \right){{\bf{e}}_x} = \left[ \begin{array}{l}
 \cos {\omega _m}\\
 \sin {\omega _m}\\
 0
 \end{array} \right],
 \end{equation}
 \begin{equation}
 {\bf{n}}\left( {{\omega _m}} \right) = {\bf{R}}\left( {{\omega _m}} \right){{\bf{e}}_y} = \left[ \begin{array}{l}
 - \sin {\omega _m}\\
  \cos {\omega _m}\\
 0
 \end{array} \right],
 \end{equation}
 where ${{\bf{e}}_x} = {\left[ {1,0,0} \right]^T}$ and ${{\bf{e}}_y} = {\left[ {0,1,0} \right]^T}$ denote the unit vectors along the $x$- and $y$-axis, respectively. Let ${{\bf{a}}_t}\left( {{x_m}} \right)$ denote the normalized incident vector from the Tx to reflector $m$, and ${{\bf{a}}_r}\left( {{x_m};{\bf r}} \right)$ denote the normalized reflection vector from reflector $m$ to location $\bf r$. Note that to enable the effective reflection, the Tx and Rx must lie on the same side of the reflector. This yields the following constraint,
 \begin{equation}\label{rotationConstraint}
 \left( {{\bf{a}}_t^T\left( {{x_m}} \right){\bf{n}}\left( {{\omega _m}} \right)} \right)\left( {{\bf{a}}_r^T\left( {{x_m};{\bf{r}}} \right){\bf{n}}\left( {{\omega _m}} \right)} \right) < 0.
 \end{equation}

 The signals reflected by two or more times by any FR are assumed to be negligible due to the high path loss. The ratio of the receive power ${P_r}\left( {{x_m},{\omega _m};{\bf{r}}} \right)$ at location $\bf r$ to the transmit power $P_t$ via reflector $m$ is given by \cite{balanis2016antenna}
 \begin{equation}\label{powerRatio}
 \frac{{{P_r}\left( {{x_m},{\omega _m};{\bf{r}}} \right)}}{{{P_t}}} = \frac{{\sigma \left( {{x_m},{\omega _m};{\bf{r}}} \right){\lambda ^2}}}{{4\pi {{\left( {4\pi {d_t}\left( {{x_m}} \right){d_r}\left( {{x_m};{\bf{r}}} \right)} \right)}^2}}},
 \end{equation}
 where $\lambda$ denotes the signal wavelength, and $\sigma \left( {{x_m},{\omega _m};{\bf{r}}} \right)$ denotes the radar cross section (RCS) of reflector $m$ observed at location $\bf r$, given by \cite{yu2023wireless}
 \begin{equation}\label{generalRCS}
 \begin{aligned}
 \sigma \left( {{x_m},{\omega _m};{\bf{r}}} \right) &= {\sigma _{\max }}\eta \left( {{x_m},{\omega _m};{\bf{r}}} \right)\times\\
 &\ \ \ \ \ {\rm{sinc}}^2\left( {\pi {{\bar L}_1}{\Delta}\left( {{x_m},{\omega _m};{\bf{r}}} \right)} \right),
 \end{aligned}
 \end{equation}
 where ${\sigma _{\max }} \triangleq 4\pi L_{1}^2L_{2}^2/{\lambda ^2}$ denotes the maximum possible RCS value, $\eta \left( {{x_m},{\omega _m};{\bf{r}} } \right)$ is a factor no greater than one, ${{\bar L}_{1}} \triangleq {L_{1}}/\lambda $ denotes the wavelength-normalized length of the reflector, $\Delta \left( {{x_m},{\omega _m};{\bf{r}}} \right) \triangleq {\left( {{{\bf{a}}_r}\left( {{x_m};{\bf{r}}} \right) - {{\bf{a}}_t}\left( {{x_m}} \right)} \right)^T}{\bf{u}}\left( {{\omega _m}} \right)$ is the projection of the deflection vector $\left( {{\bf{a}}_r}\left( {{x_m};{\bf{r}}} \right) - {{\bf{a}}_t}\left( {{x_m}} \right)\right)$ along the edge of reflector $m$, and ${\rm{sinc}}\left( x \right) \triangleq \sin \left( x \right)/x$. Moreover, the expected receive power at a small region centered at location $\bf r$ is
 \begin{equation}\label{receivedPower}
 \begin{aligned}
 {P_r}\left( {\bf{r}} \right) &= {\mathbb E}\left[{\left| {\sum\limits_{m = 1}^M {\sqrt {{P_r}\left( {{x_m},{\omega _m};{\bf{r}}} \right)} {e^{{\rm{j}}{\varphi _m}}}} } \right|^2}\right]\\
 &= \sum\limits_{m = 1}^M {{P_r}\left( {{x_m},{\omega _m};{\bf{r}}} \right)}  = \frac{{{P_t}{\lambda ^2}}}{{{{\left( {4\pi } \right)}^3}}}\sum\limits_{m = 1}^M {f\left( {{x_m},{\omega _m};{\bf{r}}} \right)},
 \end{aligned}
 \end{equation}
 where ${\varphi _m}$ denotes the signal phase of the path via reflector $m$, with ${\varphi _m}$'s being independent and identically distributed, and $f\left( {{x_m},{\omega _m};{\bf{r}}} \right) \triangleq \frac{{\sigma \left( {{x_m},{\omega _m};{\bf{r}}} \right)}}{{d_t^2\left( {{x_m}} \right)d_r^2\left( {{x_m};{\bf{r}}} \right)}}$. It is worth mentioning that different from MA that moves antenna positions to obtain a favourable channel condition, adjusting the placement position and rotation angle of each FR aims to manipulate the passive beamforming direction for the considered setup, thus achieving a higher received power.

 We aim to maximize the minimum (worst-case) expected receive power over all locations in a given target area $\cal A$, by jointly optimizing the placement positions $\left\{ {{x_m}} \right\}_{m = 1}^M$ and rotation angles $\left\{ {{\omega _m}} \right\}_{m = 1}^M$ of the $M$ FRs. The optimization problem can be formulated as (by discarding the constant terms)
 \begin{equation}\label{optimizationProblem}
 \begin{aligned}
 \left( {\rm{P1}} \right) &\  \mathop {\max }\limits_{\left\{ {{x_m},{\omega _m}} \right\}_{m = 1}^M} \ \mathop {\min }\limits_{{\bf{r}} \in {\cal A}}\ \sum\limits_{m = 1}^M {f\left( {{x_m},{\omega _m};{\bf{r}}} \right)} \\
 {\rm{s.t.}}&\ \left| {{x_m} - {x_n}} \right| \ge {d_{m,n}^{\min }},\ \forall m,n \in {\cal M},\ m \ne n,\\
 &\ \left( {{{\bf{a}}_t^T}\left( {{x_m}} \right) {\bf{n}}\left( {{\omega _m}} \right)} \right)\left( {{{\bf{a}}_r^T}\left( {{x_m};{\bf{r}}} \right)  {\bf{n}}\left( {{\omega _m}} \right)} \right) < 0,\ \forall m, {\bf r},
 \end{aligned}
 \end{equation}
 where $d_{m,n}^{\min}$ denotes the minimum distance to avoid the overlap and signal blockage between FRs $m$ and $n$. Problem (P1) is challenging to be directly solved due to the following two reasons. First, the objective function is the minimum power value across a continuous area, which is difficult to be explicitly expressed in terms of the placement positions and rotation angles of FRs. Second, problem (P1) is a non-convex optimization problem, due to the non-concave objective function and non-convex constraints.

 % >>>>>>>>>>>>>SECTIONS III -  here >>>>>>>>>>>>
 \section{Movable Reflector}\label{sectionMR}
 To gain useful insights for solving (P1), we first study the special case of MRs, by considering their fixed rotation angles ${\omega _m} = 0$, $\forall m$.

 \subsection{Single Target Location Power Enhancement}
 In this subsection, we consider the special case of single target location power enhancement, where the scenarios of single- and multi-MR are respectively studied.

 \subsubsection{Single-MR Case}
 \begin{figure}[!t]
 \centering
 \centerline{\includegraphics[width=2.69in,height=2.2in]{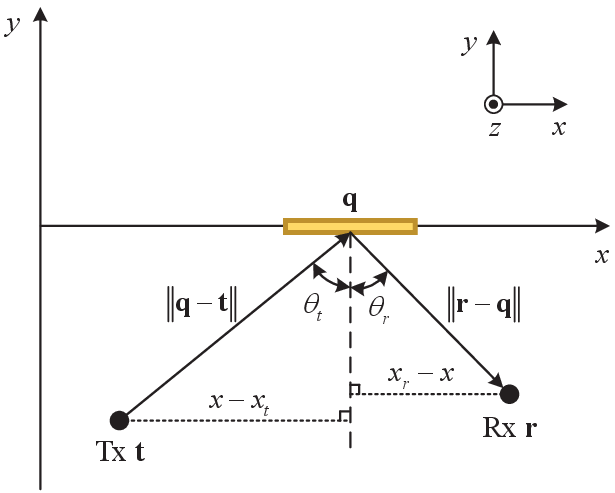}}
 \caption{Single target location power enhancement by a single-MR.}
 \label{fig:systemModelPointSingle}
 \end{figure}

 As illustrated in Fig.~\ref{fig:systemModelPointSingle}, a single-MR is deployed to assist in the communication from the Tx to Rx, where the reflector index is omitted for brevity. Besides, the location $\bf r$ is omitted in the notations for the case of single target location. The distance between the Tx and the center of the reflector is ${d_t}\left( x \right) = \left\| {{\bf{q}} - {\bf{t}}} \right\| = \sqrt {{{( {x - {x_t}} )^2}} + y_t^2} $, and that between the center of the reflector and Rx is ${d_r}\left( {x}  \right) = \left\| {{\bf{r}} - {\bf{q}}} \right\| = \sqrt {{{( {{x_r} - x} )^2}} + y_r^2} $. The normalized incident and reflection vectors are ${{\bf{a}}_t}\left( x \right) = \frac{{{\bf{q}} - {\bf{t}}}}{{\left\| {{\bf{q}} - {\bf{t}}} \right\|}} = \frac{1}{{{d_t}\left( x \right)}}{\left[ {x - {x_t}, - {y_t},0} \right]^T}$ and ${{\bf{a}}_r}\left( {x} \right) = \frac{{{\bf{r}} - {\bf{q}}}}{{\left\| {{\bf{r}} - {\bf{q}}} \right\|}} = \frac{1}{{{d_r}\left( {x} \right)}}{\left[ {{x_r} - x,{y_r},0} \right]^T}$, respectively. With the direction vector along the edge of the reflector ${\bf{u}} = {\left[ {1,0,0} \right]^T}$, we have $\Delta \left( {x} \right) = {\left( {{{\bf{a}}_r}\left( {x} \right) - {{\bf{a}}_t}\left( x \right)} \right)^T}{\bf{u}} = \frac{{{x_r} - x}}{{{d_r}\left( {x} \right)}} - \frac{{x - {x_t}}}{{{d_t}\left( x \right)}}$. Besides, the factor $\eta \left( {x} \right)$ is given by $\eta \left( {x} \right) = y_r^2/d_r^2\left( {x} \right)$ \cite{yu2023wireless}. In this case, the function $f\left( {x} \right)$ is expressed as
 \begin{equation}\label{functionfOneDimensional}
 \begin{aligned}
 &f\left( {x} \right) = \frac{{\sigma \left( {x} \right)}}{{d_t^2\left( x \right)d_r^2\left( {x} \right)}} = {\sigma _{\max }} \times  \\
 &\ \ \ \ \ \underbrace {\frac{{y_r^2}}{{d_t^2\left( x \right)d_r^4\left( {x} \right)}}}_{{f_1}\left( x \right)}\underbrace {{\rm{sinc}}^2\left( {\pi {{\bar L}_1}\left( {\frac{{{x_r} - x}}{{{d_r}\left( {x} \right)}} - \frac{{x - {x_t}}}{{{d_t}\left( x \right)}}} \right)} \right)}_{{f_2}\left( x \right)}.
 \end{aligned}
 \end{equation}
 Thus, problem (P1) is reduced to
 \begin{equation}\label{optimizationProblemSPSingle}
 \mathop {\max }\limits_x\ \  f\left( x \right).
 \end{equation}

\begin{figure}
 \centering
 \subfigure[${\bar L_1} = 10$]{
 \begin{minipage}[t]{0.5\textwidth}
 \centering
 \centerline{\includegraphics[width=3.0in,height=2.25in]{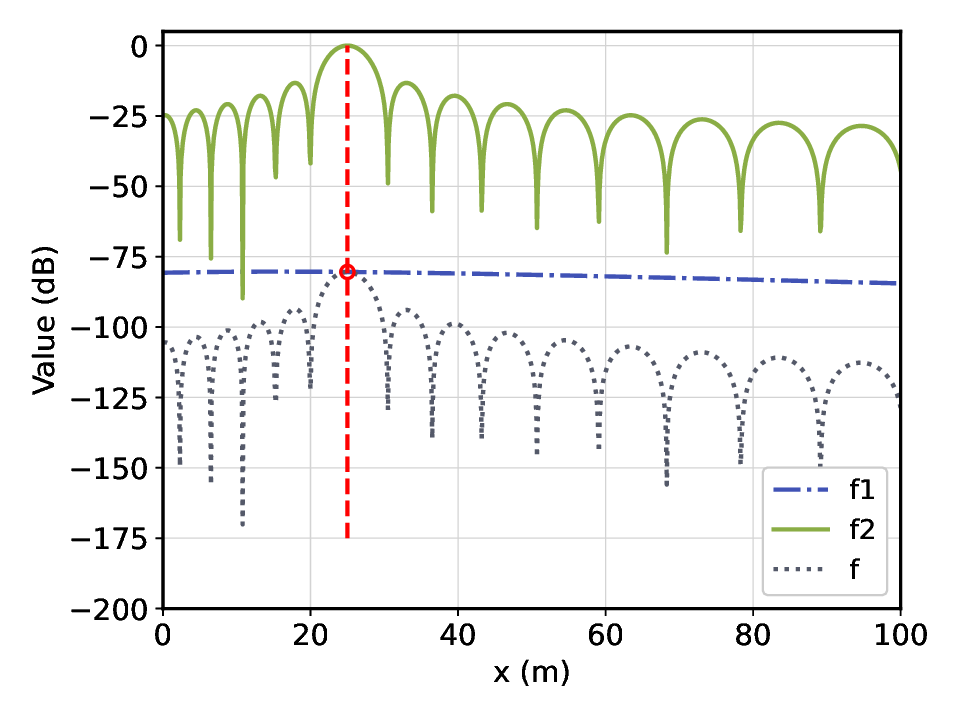}}
 \end{minipage}
 }
 \subfigure[${\bar L_1} = 0.1$ ]{
 \begin{minipage}[t]{0.5\textwidth}
 \centering
 \centerline{\includegraphics[width=3.0in,height=2.25in]{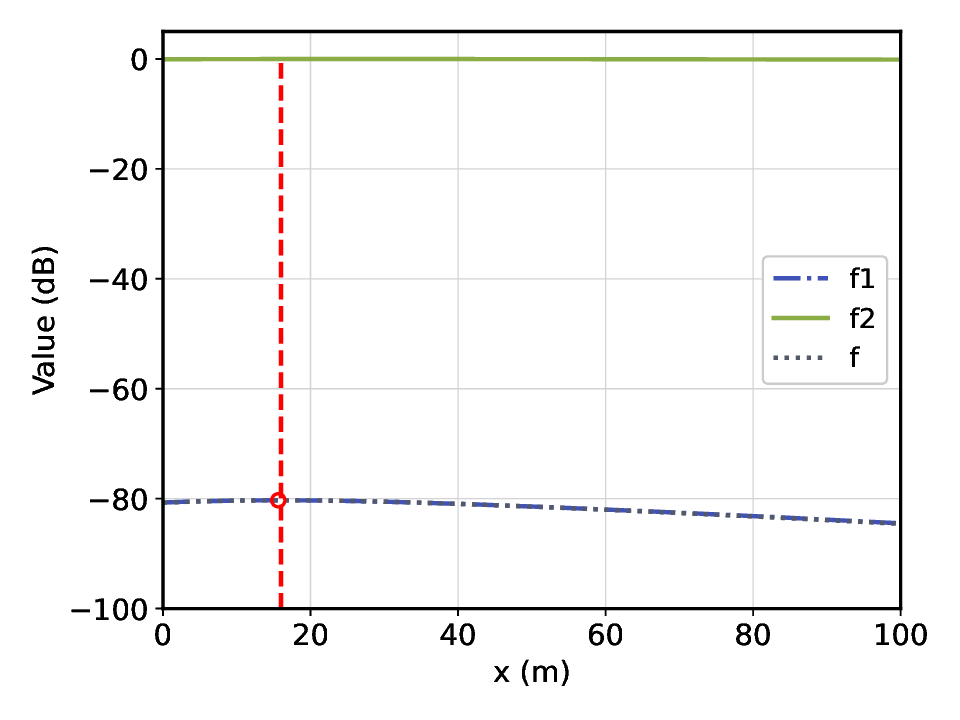}}
 \end{minipage}
 }
 \caption{Values of $f\left(x\right)$, ${{f_1}\left( {x} \right)}$ and ${{f_2}\left( {x} \right)}$ versus the placement position $x$, where the red circle and red dashed line indicate the optimal placement positions obtained by the closed-form and numerical result, respectively.}
 \label{fig:fFunctionVersusPlacement}
 \end{figure}

 It is observed from \eqref{functionfOneDimensional} that the function $f\left(x\right)$ consists of two terms ${{f_1}\left( {x} \right)}$ and ${{f_2}\left( {x} \right)}$, where ${f_1}\left( x \right)$ accounts for the factor $\eta \left( x \right)$ and the concatenated path loss from the Tx to Rx via the reflector, and ${{f_2}\left( {x} \right)}$ can be interpreted as the array factor. Fig.~\ref{fig:fFunctionVersusPlacement} illustrates the values of $f\left(x\right)$, ${{f_1}\left( {x} \right)}$ and ${{f_2}\left( {x} \right)}$ versus the placement position $x$, by considering ${\bar L_1} = 10$ and ${\bar L_1} = 0.1$, respectively. The locations of the Tx and Rx are ${\bf{t}} = {\left[ {0, - 50} \right]^T}$ m, and ${\bf{r}} = {\left[ {100, - 150} \right]^T}$ m, respectively. It is observed that for electrically large reflector with ${\bar L_1} = 10$, $f_2\left(x\right)$ composed of the sinc function exhibits significant variations as the placement position $x$ changes, as compared to $f_1 \left(x\right)$. By contrast, for electrically small reflector with ${\bar L_1} = 0.1$, the impact on $f \left(x\right)$  by $f_1 \left(x\right)$ is more significant than that by $f_2 \left(x\right)$. This is expected since the sinc function tends to be one when ${\bar L_1}$ is very small.

 \begin{proposition} \label{maximumReceivedPowerProposition}
 When ${\bar L}_1 \gg 1$, the optimal solution to \eqref{optimizationProblemSPSingle} is
 \begin{equation}\label{maximumReceivedPowerLocation}
 x^ \star  = {x_t} + \frac{{{y_t}}}{{{y_t} + {y_r}}}\left( {{x_r} - {x_t}} \right).
 \end{equation}
 \end{proposition}
 \begin{IEEEproof}
 Please refer to Appendix~\ref{proofmaximumReceivedPowerProposition}.
 \end{IEEEproof}

 A closer look at Fig.~\ref{fig:systemModelPointSingle} shows that $\sin {\theta _t} = \frac{{x - {x_t}}}{{{d_t}\left( x \right)}}$ and $\sin {\theta _r} = \frac{{{x_r} - x}}{{{d_r}\left( x \right)}}$, where $\theta_t$ and $\theta_r$ denote the incident and reflection angles, respectively. By substituting \eqref{maximumReceivedPowerLocation} into $\sin {\theta _t}$ and $\sin {\theta _r}$, we have $\sin {\theta _t} = \sin {\theta _r} = \frac{{{x_r} - {x_t}}}{{\sqrt {{{\left( {{x_r} - {x_t}} \right)}^2} + {{\left( {{y_t} + {y_r}} \right)}^2}} }}$. This implies that for electrically large reflector, the reflector should be placed at the specular reflection point with ${\theta _r} = {\theta _t}$.

 On the other hand, for electrically small reflector, the placement position variation has a negligible impact on $f_2\left(x\right)$, which is expected since the reflector tends to be isotropic. In this case, the placement position $x$ is designed to maximize $f_1\left(x\right)$. The problem can be equivalently formulated as
 \begin{equation}\label{sincMaximizationProblemSmallSTL}
 \begin{aligned}
 \mathop {{\rm{min}}}\limits_{x} &\ \  \left( {{{\left( {x - {x_t}} \right)}^2} + y_t^2} \right){\left( {{{\left( {{x_r} - x} \right)}^2} + y_r^2} \right)^2}.
 \end{aligned}
 \end{equation}

 \begin{lemma} \label{maximumReceivedPowerSmalllemma}
 The optimal solution to \eqref{sincMaximizationProblemSmallSTL} can be obtained by solving the following equation
 \begin{equation}\label{smallDerivationEquation}
 {a_c}{x^3} + {b_c}{x^2} + {c_c}x + {d_c} = 0,
 \end{equation}
 where ${a_c} = 3$, ${b_c} =  - \left( {5{x_t} + 4{x_r}} \right)$, ${c_c} = \left( {x_r^2 + y_r^2} \right) + 2\left( {x_t^2 + y_t^2} \right) + 6{x_r}{x_t}$, and ${d_c} =  - \left[ {\left( {x_r^2 + y_r^2} \right){x_t} + } \right.$ $\left. {2\left( {x_t^2 + y_t^2} \right){x_r}} \right]$, respectively.
 \end{lemma}
 \begin{IEEEproof}
 Please refer to Appendix~\ref{proofOfmaximumReceivedPowerSmalllemma}.
 \end{IEEEproof}

 Since \eqref{smallDerivationEquation} is a cubic equation, it can be solved in closed-form. Then, the optimal placement position corresponds to the solution with the minimum objective function.

 It is observed from Fig.~\ref{fig:fFunctionVersusPlacement} that the optimal placement positions obtained by the closed-form well match with those obtained by numerical results for both electrically large and small reflectors. Moreover, in contrast to the electrically small reflector where only a slight performance variation is observed as the placement position changes, the reflector placement position optimization brings considerable performance gain of several tens of dB for the electrically large reflector. Such a significant gain motivates us to focus on the electrically large reflector in the following.

 \subsubsection{Multi-MR Case}
 Next, we consider the case of single target location power enhancement assisted by multiple MRs. In this case, problem (P1) is reduced to
 \begin{equation}\label{optimizationProblemSingleLocationMulti}
 \begin{aligned}
 \mathop {\max }\limits_{\left\{ {{x_m}} \right\}_{m = 1}^M} &\ \  \sum\limits_{m = 1}^M {f\left( {{x_m}} \right)}  \\
 {\rm{s.t.}}&\ \left| {{x_m} - {x_n}} \right| \ge {d_{m,n}^{\min }},\ \forall m,n \in {\cal M},\ m \ne n,
 \end{aligned}
 \end{equation}
 where ${d_{m,n}^{\min }} = {L_1}$ corresponds to the minimum distance to avoid the overlap of MRs.

 Before solving \eqref{optimizationProblemSingleLocationMulti}, we first study the property of the function $g\left( \Delta  \right) \triangleq {\rm{sinc}}^2 \left( {\pi L\Delta } \right)$. The null-to-null beam width can be obtained by letting $\pi L\Delta  =  \pm \pi $, which is given by $2/L$. The main lobe width is then defined as half of the null-to-null beam width, i.e., the range of main lobe is $\left[ { - \frac{1}{2L},\frac{1}{2L}} \right]$, and $G \triangleq g \left( {\frac{1}{{2L}}} \right) \approx 0.4$ is the value at the endpoint of the beamforming main lobe. Moreover, the value of $g\left(\Delta \right)$ monotonically decreases within the main lobe as $\left| \Delta  \right|$ increases. Thus, for electrically large reflectors, the reflector should be placed such that the corresponding $\left| \Delta  \right|$ is as small as possible.

 Motivated by the above discussions, the placement positions of reflectors are designed such that the Rx is located within their beamforming main lobes. Specifically, for ${f_2}\left(x \right)$, the placement positions $x$ corresponding to the boundary points of the beamforming main lobe can be obtained by letting $\frac{{{x_r} - x}}{{{d_r}\left( x \right)}} - \frac{{x - {x_t}}}{{{d_t}\left( x \right)}}=  \pm \frac{1}{{2{{\bar L}_1}}}$, and the solution can be found numerically, denoted as ${x_{bl}}$ and ${x_{br}}$, respectively. It is worth mentioning that the projection $\Delta \left(x\right)$ will decrease as the reflector moves from the Tx to Rx. Thus, as $x$ increases, the placement position will first reach the right boundary point of the beamforming main lobe, and then the left boundary point of the beamforming main lobe, i.e., ${x_{br}} < {x_{bl}}$. In particular, when the placement position is out of $\left[ {{x_{br}},{x_{bl}}} \right]$, the reflector contributes little to the receive power, and thus this case is not considered here. To ensure that  $\left| {\Delta \left( {{x_m}} \right)} \right|$ of reflector $m$ is as small as possible, the reflector $m$ should be placed close to $x^{\star}$ given in \eqref{maximumReceivedPowerLocation}. Thus, the actual number of reflectors to be deployed is
 \begin{equation}
 \tilde M = \left\lfloor {\frac{{{x^ \star } - {x_{br}}}}{{{L_1}}}} \right\rfloor  + \left\lfloor {\frac{{{x_{bl}} - {x^ \star }}}{{{L_1}}}} \right\rfloor  + 1,
 \end{equation}
 and the placement position of reflector $m$, $1 \le m \le {\tilde M}$, is
 \begin{equation}\label{multiReflectorPlacementSTL}
 x_m^ \star  = {x^ \star } - \left\lfloor {\frac{{{x^ \star } - {x_{br}}}}{{{L_1}}}} \right\rfloor {L_1} + \left( {m - 1} \right){L_1}.
 \end{equation}

\subsection{Area Coverage Enhancement}
 In this subsection, we consider the case of area coverage enhancement.

\subsubsection{Single-MR Case}\label{subsubsectionSingleMRArea}
 For the target area illustrated in Fig.~\ref{fig:systemModel}, let ${{\bf{r}}_{ul}} = {\left[ {{x_c} - {D_x}/2,{y_c} + {D_y}/2} \right]^T}$, ${{\bf{r}}_{ll}} = {\left[ {{x_c} - {D_x}/2,{y_c} - {D_y}/2} \right]^T}$, ${{\bf{r}}_{ur}} = {\left[ {{x_c} + {D_x}/2,{y_c} + {D_y}/2} \right]^T}$, and ${{\bf{r}}_{lr}} = {\left[ {{x_c} +{D_x}/2,{y_c} - {D_y}/2} \right]^T}$ denote the upper left, lower left, upper right and lower right corners of $\cal A$, respectively. In the case of single-MR aided area coverage enhancement, problem (P1) becomes
 \begin{equation}\label{originalProblemAreaSingle}
 \mathop {\max }\limits_x\ \  \mathop {\min }\limits_{{\bf{r}} \in {\cal A}}\ {\sigma _{\max }}{f_1}\left( {x;{\bf{r}}} \right){f_2}\left( {x;{\bf{r}}} \right).
 \end{equation}

 For any given reflector placement position $x$, let ${\Delta _{\max }}\left( x \right)$ and ${\Delta _{\min }}\left( x \right)$ denote the maximum and minimum projection values in the target area $\cal A$, given by
 \begin{equation}\label{areaMaximumDeviation}
 {\Delta _{\max }}\left( x \right) \triangleq \mathop {\max }\limits_{{\bf{r}} \in {\cal A}} \frac{{{x_r} - x}}{{\sqrt {{{\left( {{x_r} - x} \right)}^2} + y_r^2} }} - \frac{{x - {x_t}}}{{\sqrt {{{\left( {x - {x_t}} \right)}^2} + y_t^2} }},
 \end{equation}
 \begin{equation}\label{areaMinimumDeviation}
 {\Delta _{\min }}\left( x \right) \triangleq \mathop {\min }\limits_{{\bf{r}} \in {\cal A}} \frac{{{x_r} - x}}{{\sqrt {{{\left( {{x_r} - x} \right)}^2} + y_r^2} }} - \frac{{x - {x_t}}}{{\sqrt {{{\left( {x - {x_t}} \right)}^2} + y_t^2} }}.
 \end{equation}

 For electrically large reflector, problem \eqref{originalProblemAreaSingle} is approximated as
 \begin{equation}\label{reducedProblemSingleMRArea}
 \mathop {\max }\limits_x\ \  \mathop {\min }\limits_{{\Delta _{\min }}\left( x \right) \le \Delta  \le {\Delta _{\max }}\left( x \right)} {f_2}\left( {x;\Delta } \right) \triangleq {\rm{sinc}}^2\left( {\pi {{\bar L}_1}\Delta } \right).
 \end{equation}

 The null points of ${f_2}\left( {x;\Delta } \right)$ can be obtained by letting $\pi {{\bar L}_1}\Delta  =  \pm z\pi $, $z \in {\cal N}_{+}$, i.e., $\Delta  =  \pm z/{{\bar L}_1}$. For any given placement position $x$, if there exist one or more null points in the range $\left[ {{\Delta _{\min }}\left( x \right),{\Delta _{\max }}\left( x \right)} \right]$, the minimum value within the target area is zero. Otherwise, the minimum value in the target area occurs at ${{\Delta _{\min }}\left( x \right)}$ or ${{\Delta _{\max }}\left( x \right)}$. Thus, we have
 \begin{equation}\label{minimumReflectionGainArea}
 \begin{aligned}
 &\mathop {\min }\limits_{{\Delta _{\min }}\left( x \right) \le \Delta  \le {\Delta _{\max }}\left( x \right)} {\rm{sin}}{{\rm{c}}^2}\left( {\pi {{\bar L}_1}\Delta } \right) = \\
 &\left\{ \begin{split}
 &0,\ \ {\rm{if }}\ \exists z \in {{\cal N}_+ },\ {\rm s.t.}\ \pm z/{{\bar L}_1} \in \left[ {{\Delta _{\min }}\left( x \right),{\Delta _{\max }}\left( x \right)} \right],\\
 &\min \left\{ {{\rm{sin}}{{\rm{c}}^2}\left( {\pi {{\bar L}_1}{\Delta _{\min }}\left( x \right)} \right),{\rm{sin}}{{\rm{c}}^2}\left( {\pi {{\bar L}_1}{\Delta _{\max }}\left( x \right)} \right)} \right\},\\
 &\ \ \ \ \ {\rm otherwise}.
 \end{split} \right.
 \end{aligned}
 \end{equation}

 With \eqref{minimumReflectionGainArea}, the optimal solution to \eqref{reducedProblemSingleMRArea} can be effectively obtained via the one-dimensional search. Furthermore, the search region can be reduced by noting that the specular reflection direction should point towards the target area. Specifically, as the reflector moves towards the target area, the specular reflection direction first points towards the lower left corner of $\cal A$, i.e., ${{\bf{r}}_{ll}}$, and finally points towards the upper right corner, i.e., ${{\bf{r}}_{ur}}$. The minimum and maximum placement positions are then determined by regarding the lower left and upper right corners as the single target location, respectively. With Proposition \ref{maximumReceivedPowerProposition}, we have
 \begin{equation}\label{lowerBoundSearch}
 {x_{\rm lower}} = {x_t} + \frac{{{y_t}}}{{{y_t} + {y_c} - \frac{{{D_y}}}{2}}}\left( {{x_c} - \frac{{{D_x}}}{2} - {x_t}} \right),
 \end{equation}
 \begin{equation}\label{upperBoundSearch}
 {x_{\rm upper}} = {x_t} + \frac{{{y_t}}}{{{y_t} + {y_c} + \frac{{{D_y}}}{2}}}\left( {{x_c} + \frac{{{D_x}}}{2} - {x_t}} \right).
 \end{equation}
 Thus, the search region of the placement is reduced to $\left[ {{x_{\rm lower}},{x_{\rm upper}}} \right]$. The main procedures for solving problem \eqref{reducedProblemSingleMRArea} are summarized in Algorithm~\ref{alg1}.

 \begin{algorithm}[t]
 \caption{Reflector Placement Position Optimization for \eqref{reducedProblemSingleMRArea}}
 \label{alg1}
 \begin{algorithmic}[1]
 \STATE \textbf{Initialization:} Set the region of the reflector placement position $x \in \left[ {{x_{{\rm{lower}}}},{x_{{\rm{upper}}}}} \right]$ based on \eqref{lowerBoundSearch} and \eqref{upperBoundSearch}.
 \STATE For any candidate placement position $x$, calculate the maximum projection value ${\Delta _{\max }}\left( x \right)$ and the minimum projection value ${\Delta _{\min }}\left( x \right)$ based on \eqref{areaMaximumDeviation} and \eqref{areaMinimumDeviation}, respectively.
 \STATE Obtain the minimum value of ${\rm{sinc}}^2\left( {\pi {{\bar L}_1}\Delta } \right)$ within $\cal A$ based on \eqref{minimumReflectionGainArea}.
 \STATE Choose the $x$ that gives the maximum value in \eqref{reducedProblemSingleMRArea} as the optimal placement position.
 \end{algorithmic}
 \end{algorithm}

\subsubsection{Multi-MR Case}\label{subsubsectionMultiMRArea}
 Furthermore, we consider the case of area coverage enhancement with multiple MRs. The receive power at any location ${\bf r}$ within $\cal A$ can be expressed as
 \begin{equation}\label{receivedPowerMultiReflector}
 {P_r}\left( {\bf{r}} \right) = \sum\limits_{m = 1}^M {\frac{{{P_t}{\lambda ^2}}}{{{{\left( {4\pi } \right)}^3}}}{\sigma _{\max }}{f_1}\left( {{x_m};{\bf{r}}} \right)} {f_2}\left( {{x_m};{\bf{r}}} \right).
 \end{equation}

 For electrically large reflector, problem (P1) is approximated as
 \begin{equation}\label{problemAreaMultiMR}
 \begin{aligned}
 \mathop {\max }\limits_{\left\{ {{x_m}} \right\}_{m = 1}^M} &\ \   \mathop {\min }\limits_{{\bf{r}} \in {\cal A}}\ \ \sum\limits_{m = 1}^M {{f_2}\left( {{x_m};{\bf{r}}} \right)} \\
 {\rm{s.t.}}&\ \ \left| {{x_m} - {x_n}} \right| \ge {L_1},\ \forall m,n \in {\cal M},\ m \ne n.
 \end{aligned}
 \end{equation}
 Even for the simplified problem \eqref{problemAreaMultiMR}, it is still difficult to be directly solved. To tackle this issue, we propose a sequential placement algorithm, where any location of the target area can be covered by the beamforming main lobe of one reflector.

 Let $x_m^ \star $ denote the optimized placement position of MR $m$, and ${{\bf{r}}_{m,l}}$ and ${{\bf{r}}_{m,r}}$ denote the locations corresponding to left and right endpoints of the beamforming main lobe with respect to (w.r.t.) MR $m$, respectively, as illustrated in Fig.~\ref{illustrationOFCoverageArea} in the appendix. For the proposed sequential placement algorithm, the parameters are determined in the following order,
 \begin{equation}
 {{\bf{r}}_{1,r}} \to x_1^ \star  \to {{\bf{r}}_{1,l}}\left( {{{\bf{r}}_{2,r}}} \right) \to x_2^ \star  \to {{\bf{r}}_{2,l}}\left( {{{\bf{r}}_{3,r}}} \right) \to  \cdots.
 \end{equation}

 Specifically, by choosing the upper right corner ${\bf r}_{ur}$ as the location corresponding to the right endpoint of the beamforming main lobe w.r.t. the first MR, we have ${{\bf{r}}_{1,r}} = {\bf r}_{ur}$. Then, for MR $m$, the placement position $x_m$ is designed such that ${\Delta}\left( {{x_m};{{\bf{r}}_{m,r}}} \right) = \frac{1}{{2{{\bar L}_1}}}$, i.e.,
 \begin{equation}\label{rightEndpointEqu}
 \frac{{{x_{m,r}} - {x_m}}}{{\sqrt {{{\left( {{x_{m,r}} - {x_m}} \right)}^2} + y_{m,r}^2} }} - \frac{{{x_m} - {x_t}}}{{\sqrt {{{\left( {{x_m} - {x_t}} \right)}^2} + y_t^2} }} = \frac{1}{{2{{\bar L}_1}}},
 \end{equation}
 where ${{x_{m,r}}}$ and ${{y_{m,r}}}$ denote the $x$- and $y$-coordinate of ${{{\bf{r}}_{m,r}}}$, respectively, and the solution, $x_m^ \star $, can be obtained numerically. With $x_m^{\star}$, the projection of ${\bf{r}}_{ll}$ w.r.t. MR $m$, i.e., ${\Delta}\left( {x_m^ \star ;{{\bf{r}}_{ll}}} \right)$, is obtained. When ${\Delta}\left( {x_m^ \star ;{{\bf{r}}_{ll}}} \right) \in \left[ { - \frac{1}{{2{{\bar L}_1}}},\frac{1}{{2{{\bar L}_1}}}} \right]$, the lower left corner can be covered by the beamforming main lobe of MR $m$, and the sequential placement is ended. Otherwise, the next MR is then deployed. To this end, the location corresponding to the left endpoint of the beamforming main lobe w.r.t. MR $m$ needs to be determined, as shown in the following lemma.

 \begin{lemma}\label{leftEndPointlemma}
 A location corresponding to the left endpoint of the beamforming main lobe w.r.t. MR $m$ can be expressed as
 \begin{equation}\label{leftEndpointExpression}
 \begin{aligned}
 &{{\bf{r}}_{m,l}} = {\left[ {{x_{m,l}},{y_{m,l}}} \right]^T} = \\
 &\left\{ \begin{split}
 &{\left[ {{x_r},{y_c} + \frac{{{D_y}}}{2}} \right]^T},\ {\rm if}\ {\Delta}\left( {x_m^ \star ;{{\bf{r}}_{ul}}} \right) <  - \frac{1}{{2{{\bar L}_1}}},\\
 &{\left[ {{x_c} - \frac{{{D_x}}}{2},{y_r}} \right]^T},\ {\rm if}\ {\Delta}\left( {x_m^ \star ;{{\bf{r}}_{ul}}} \right) \ge  - \frac{1}{{2{{\bar L}_1}}}\ {\rm and} \\
 &\ \ \ \ \ \ \ \ \ \ \ \ \ \ \ \ \ \ \ \ \ \ \ \ \ \ \ \ \ \ {\Delta}\left( {x_m^ \star ;{{\bf{r}}_{ll}}} \right) <  - \frac{1}{{2{{\bar L}_1}}},
 \end{split} \right.
 \end{aligned}
 \end{equation}
 where $x_r$ and $y_r$ are the values within $\left( {{x_c} - {D_x}/2},\right.$ $\left.{{x_c} + {D_x}/2} \right)$ and $\left( {{y_c} - {D_y}/2,{y_c} + {D_y}/2} \right)$, respectively, and ${\Delta}\left( {x_m^ \star ;{{\bf{r}}_{ul}}} \right)$ denotes the projection of the upper left corner ${\bf r}_{ul}$ w.r.t. MR $m$.
 \end{lemma}

 \begin{IEEEproof}
 Please refer to Appendix~\ref{proofOfleftEndPointlemma}.
 \end{IEEEproof}

 Furthermore, the unknown parameter in Lemma~\ref{leftEndPointlemma} can be obtained by solving the equation ${\Delta}\left( {x_m^ \star ,{{\bf{r}}_{m,l}}} \right) =  - \frac{1}{{2{{\bar L}_1}}}$, i.e.,
 \begin{equation}\label{leftEndpointEqu}
 \frac{{{x_{m,l}} - x_m^ \star }}{{\sqrt {{{\left( {{x_{m,l}} - x_m^ \star } \right)}^2} + y_{m,l}^2} }} - \frac{{x_m^ \star  - {x_t}}}{{\sqrt {{{\left( {x_m^ \star  - {x_t}} \right)}^2} + y_t^2} }} =  - \frac{1}{{2{{\bar L}_1}}}.
 \end{equation}

 Subsequently, the $\left(m +1\right)$-th MR is placed such that ${{\bf{r}}_{m,l}}$ is covered by the right endpoint of the beamforming main lobe w.r.t. MR $m+1$, i.e., ${{\bf{r}}_{m+1,r}} = {{\bf{r}}_{m,l}}$. Similarly, the parameters $x_{m+1}^ \star $ and ${{\bf{r}}_{m+1,l}}$ can be obtained. Note that when $\left| {x_{m+1}^ \star  - x_m^ \star } \right| \ge {L_1}$, the placement strategy constitutes a feasible solution. Otherwise, one cannot find a solution to ensure that any location of the target area is covered by the beamforming main lobe of one MR.

 Finally, the sequential placement is completed until the lower left corner ${\bf{r}}_{ll}$ is located within the beamforming main lobe of one reflector. The main procedures of the proposed sequential placement algorithm are summarized in Algorithm~\ref{alg2}.

 \begin{algorithm}[t]
 \caption{Sequential Placement Algorithm for Solving \eqref{problemAreaMultiMR}}
 \label{alg2}
 \begin{algorithmic}[1]
 \STATE Initialize ${{\bf{r}}_{1,r}} = {{\bf{r}}_{ur}}$, and let $m=1$.
 \REPEAT
 \STATE For given ${{\bf r}_{m,r}}$, solve the equation \eqref{rightEndpointEqu}, and denote the solution as $x_m^ \star$.
 \STATE For given $x_m^ \star$, solve the equation \eqref{leftEndpointEqu}, and denote the solution as ${\bf r}_{m,l}$.
 \STATE ${{\bf{r}}_{m + 1,r}} = {{\bf{r}}_{m,l}}$.
 \STATE Update $m = m+1$.
 \UNTIL ${\bf r}_{ll}$ is covered by the beamforming main lobe of one MR.
 \end{algorithmic}
 \end{algorithm}

%% >>>>>>>>>>>>>SECTIONS IV  -  here >>>>>>>>>>>>
 \section{Flexible Reflector}\label{sectionFR}
 In this section, we study the general FR-aided coverage enhancement.

 \subsection{Single Target Location Power Enhancement}
 \subsubsection{Single-FR Case}\label{subsubSectionSingleFRSL}
 For the single target location power enhancement by a single-FR, we have
 \begin{equation}
 \begin{aligned}
 f\left( {x,\omega } \right) &= \frac{{\sigma \left( {x,\omega } \right)}}{{d_t^2\left( x \right)d_r^2\left( x \right)}} = {\sigma _{\max }} \times\\
 &\underbrace {\frac{{\eta \left( {x,\omega } \right)}}{{d_t^2\left( x \right)d_r^2\left( x \right)}}}_{{f_1}\left( {x,\omega } \right)}\underbrace {{\rm{sin}}{{\rm{c}}^2}\left( {\pi {{\bar L}_1}\Delta \left( {x,\omega } \right)} \right)}_{{f_2}\left( {x,\omega } \right)},
 \end{aligned}
 \end{equation}
 and problem (P1) is reduced to
 \begin{equation}\label{optimizationProblemSingleLocationSFR}
 \begin{aligned}
 \mathop {\max }\limits_{x,\omega }&\ \  f\left( {x,\omega } \right)\\
 {\rm{s.t.}}&\ \left( {{{\bf{a}}_t^T}\left( x \right) {\bf{n}}\left( \omega  \right)} \right)\left( {{{\bf{a}}_r^T}\left( {x } \right)  {\bf{n}}\left( \omega  \right)} \right) < 0,
 \end{aligned}
 \end{equation}
 where the notation of location ${\bf r}$ is omitted.

 To tackle the non-convex problem \eqref{optimizationProblemSingleLocationSFR}, for any given placement position $x$, the rotation angle is optimized to maximize the following problem
 \begin{equation}\label{optimizationProblemSingleLocationSFRS1}
 \begin{aligned}
 \mathop {\max }\limits_{\omega }&\ \ {\rm{sinc}}^2\left( {\pi {{\bar L}_1}\Delta \left( {x,\omega } \right)} \right)\\
 {\rm{s.t.}}&\ \left( {{{\bf{a}}_t^T}\left( x \right)  {\bf{n}}\left( \omega  \right)} \right)\left( {{{\bf{a}}_r^T}\left( {x } \right)  {\bf{n}}\left( \omega  \right)} \right) < 0.
 \end{aligned}
 \end{equation}

 Then, by substituting the obtained solution, denoted as ${\omega}^{\star}\left(x\right)$, into the objective function of \eqref{optimizationProblemSingleLocationSFR}, the placement position optimization yields the following problem
 \begin{equation}\label{optimizationProblemSingleLocationSFRS2}
 \mathop {\max }\limits_{x }\ \ f\left( {x,{{\omega}^{\star}\left(x\right)}} \right).
 \end{equation}

 \begin{proposition}\label{rotationAngleproposition}
 An optimal solution to problem \eqref{optimizationProblemSingleLocationSFRS1} is
 \begin{equation}\label{zeroDeviationRotationAngle}
 \omega^{\star} \left( x \right) = \arctan \left( { - \frac{{\frac{1}{{{d_r}\left( x \right)}}\left( {{x_r} - x} \right) - \frac{1}{{{d_t}\left( x \right)}}\left( {x - {x_t}} \right)}}{{\frac{1}{{{d_r}\left( x \right)}}{y_r} + \frac{1}{{{d_t}\left( x \right)}}{y_t}}}} \right).
 \end{equation}
 \end{proposition}
 \begin{IEEEproof}
 Please refer to Appendix~\ref{proofOfrotationAngleproposition}.
 \end{IEEEproof}

 Note that after rotation by angle $\omega^{\star} \left( x \right)$, the relationship between the new incident angle ${\tilde {\theta} _t}$ and original incident angle ${\theta _t}$ is ${\tilde {\theta} _t} = {\theta _t} + {\omega ^ \star }\left( x \right)$, with $\cos ( {{{\tilde \theta }_t}} ) = \frac{{ - {{\bf{a}}_t^T}\left( x \right)  \left( { - {\bf{n}}\left( {{\omega ^ \star }\left( x \right)} \right)} \right)}}{{\left\| {{{\bf{a}}_t}\left( x \right)} \right\|\left\| {{\bf{n}}\left( {{\omega ^ \star }\left( x \right)} \right)} \right\|}}  =  - \frac{{{{\bf{a}}_t^T}\left( x \right)  {{\bf{a}}_r}\left( x \right) - 1}}{{\left\| {{{\bf{a}}_r}\left( x \right) - {{\bf{a}}_t}\left( x \right)} \right\|}}$. Besides, the relationship between the new reflection angle ${\tilde {\theta }_r}$ and original reflection angle ${\theta _r}$ is ${\tilde {\theta }_r} = {\theta _r} - {\omega ^ \star }\left( x \right)$, with $\cos ({{\tilde \theta }_r}) = \frac{{{{\bf{a}}_r^T}\left( x \right)  \left( { - {\bf{n}}\left( {{\omega ^ \star }\left( x \right)} \right)} \right)}}{{\left\| {{{\bf{a}}_r}\left( x \right)} \right\|\left\| {{\bf{n}}\left( {{\omega ^ \star }\left( x \right)} \right)} \right\|}} =  - \frac{{{{\bf{a}}_t^T}\left( x \right) {{\bf{a}}_r}\left( x \right) - 1}}{{\left\| {{{\bf{a}}_r}\left( x \right) - {{\bf{a}}_t}\left( x \right)} \right\|}}$. Thus, we have $\cos ({{\tilde \theta }_t}) = \cos ({{\tilde \theta }_r})$, i.e., simple rotation achieves the manipulation of specular direction pointing towards the Rx.

 With the rotation angle \eqref{zeroDeviationRotationAngle}, and after some manipulations, the objective function in \eqref{optimizationProblemSingleLocationSFR} is reduced to
 \begin{equation}\label{objectiveFunctionOptimalRotation}
 \begin{aligned}
 &f\left( {x,{\omega ^ \star }\left( x \right)} \right) = {\sigma _{\max }}{f_1}\left( {x,{\omega ^ \star }\left( x \right)} \right) = {\sigma _{\max }}\frac{{\eta \left( {x,{\omega ^ \star }\left( x \right)} \right)}}{{d_t^2\left( x \right)d_r^2\left( x \right)}}\\
 &\ ={\sigma _{\max }} \frac{{{{\cos }^2}\left( {{\theta _r} - {\omega ^ \star }\left( x \right)} \right)}}{{d_t^2\left( x \right)d_r^2\left( x \right)}}
 ={\sigma _{\max }} \frac{{{{\left\| {{{\bf{a}}_r}\left( x \right) - {{\bf{a}}_t}\left( x \right)} \right\|}^2}}}{{4d_t^2\left( x \right)d_r^2\left( x \right)}}.
 \end{aligned}
 \end{equation}
 Thus, problem \eqref{optimizationProblemSingleLocationSFRS2} can be equivalently expressed as
 \begin{equation}\label{newProblemSingleLocationSFRS2}
 \mathop {\max }\limits_{x }\ \ \frac{{{{\left\| {{{\bf{a}}_r}\left( x \right) - {{\bf{a}}_t}\left( x \right)} \right\|}^2}}}{{4d_t^2\left( x \right)d_r^2\left( x \right)}}.
 \end{equation}

 Since the resulting objective function is a highly complicated function w.r.t. placement position $x$, the one-dimensional search is applied to efficiently obtain the optimal FR placement position $x_{\rm FR}^{\star}$.

 \subsubsection{Multi-FR Case}\label{subsubsectionMultiFRSL}
 For multi-FR aided single target location power enhancement, problem (P1) is reduced to
 \begin{equation}\label{problemSPMultiFR}
 \begin{aligned}
 &\mathop {\max }\limits_{\left\{ {{x_m},{\omega _m}} \right\}_{m = 1}^M} \ \ \sum\limits_{m = 1}^M {{f}\left( {{x_m},{\omega _m}} \right)} \\
 {\rm{s.t.}}&\ \left| {{x_m} - {x_n}} \right| \ge {d_{m,n}^{\min }},\ \forall m,n \in {\cal M},\ m \ne n,\\
 &\ \left( {{{\bf{a}}_t^T}\left( {{x_m}} \right)  {\bf{n}}\left( {{\omega _m}} \right)} \right)\left( {{{\bf{a}}_r^T}\left( {{x_m}} \right)  {\bf{n}}\left( {{\omega _m}} \right)} \right) < 0,\ \forall m, {\bf r}.
 \end{aligned}
 \end{equation}

 To solve this problem, we first analyze the minimum distance to avoid the signal blockage, given in the following lemma.
 \begin{lemma}\label{distanceAntiBlockageFRlemma}
 For any given placement position and rotation angle pair $\left( {x,\omega } \right)$, the minimum distance to avoid the signal blockage is
 \begin{equation}\label{distanceAntiBlockageFR}
 {d}\left( {x, \omega}\right) = \max \left( {{d_i}\left( {x,\omega } \right),{d_r}\left( {x,\omega } \right)} \right),
 \end{equation}
 where ${d_i}\left( {x,\omega } \right) \triangleq \frac{{{L_1}}}{2}\left( {\cos \omega  - \tan {\theta _t}\sin \omega  + \frac{1}{{\cos {\theta _t}}}} \right)$ and ${d_r}\left( {x,\omega } \right) \triangleq \frac{{{L_1}}}{2}\left( {\cos \omega  + \tan {\theta _r}\sin \omega  + \frac{1}{{\cos {\theta _r}}}} \right)$, respectively.
 \end{lemma}

 \begin{IEEEproof}
 Please refer to Appendix~\ref{proofOfDistanceAntiBlockageFR}.
 \end{IEEEproof}

 Then, the deployment of multi-FR will utilize the result given in Section~\ref{subsubSectionSingleFRSL}, i.e., FRs are placed close to $x_{\rm FR}^{\star}$, without giving rise to the signal blockage. Meanwhile, the rotation angle of each FR is adjusted such that the reflective beamforming points towards the Rx. Let $\Omega $ denote the set of candidate placement positions for the deployed FRs. The main procedures are summarized in Algorithm~\ref{alg3}.

 \begin{algorithm}[t]
 \caption{Multi-FR Design for Solving \eqref{problemSPMultiFR}}
 \label{alg3}
 \begin{algorithmic}[1]
 \STATE Initialize $\Omega  = \emptyset$, $x_1^ +  = x_1^ - = x_{{\rm{FR}}}^ \star$, and let $i = j =1$.
 \STATE \textbf{while} $i \le \left\lceil {\left( {M + 1} \right)/2} \right\rceil$
 \STATE \quad $\Omega  = \Omega  \cup {x_i^{+}}$.
 \STATE \quad Calculate ${\omega ^ \star }\left( {x_i^ + } \right)$ based on \eqref{zeroDeviationRotationAngle}.
 \STATE \quad $x_{i + 1}^ +  = x_i^ +  + d\left( {x_i^ + ,{\omega ^ \star }\left( {x_i^ + } \right)} \right)$.
 \STATE \quad Update $i = i+1$.
 \STATE \textbf{end while}
 \STATE \textbf{while} $j \le \left\lceil {\left( {M + 1} \right)/2} \right\rceil$
 \STATE \quad $\Omega  = \Omega  \cup {x_j^{-}}$.
 \STATE \quad Calculate ${\omega ^ \star }\left( {x_j^ - } \right)$ based on \eqref{zeroDeviationRotationAngle}.
 \STATE \quad $x_{j + 1}^ -  = x_j^ -  - d\left( {x_j^ - ,{\omega ^ \star }\left( {x_j^ - } \right)} \right)$.
 \STATE \quad Update $j = j+1$.
 \STATE \textbf{end while}
 \STATE $\Omega  = \Omega \backslash x_{{\rm{FR}}}^ \star$.
 \STATE Calculate the objective function values \eqref{objectiveFunctionOptimalRotation} of all candidate placement positions within $\Omega$.
 \STATE Select the placement positions with $M$ largest values as the optimized FR placement positions.
 \end{algorithmic}
 \end{algorithm}

 \subsection{Area Coverage Enhancement}
 In this subsection, we consider the FR-aided area coverage enhancement.
 \subsubsection{Single-FR Case}\label{subsubsectionSingleFRArea}
 For the single-FR aided area coverage enhancement, problem (P1) becomes
 \begin{equation}\label{optimizationProblemSingleFRArea}
 \begin{aligned}
 \mathop {\max }\limits_{x,\omega }&\ \ \mathop {\min } \limits_{{\bf{r}} \in {\cal A}} \ \ f\left( {x,\omega ;{\bf{r}}} \right)\\
 {\rm{s.t.}}&\ \left( {{{\bf{a}}_t^T}\left( x \right)  {\bf{n}}\left( \omega  \right)} \right)\left( {{{\bf{a}}_r^T}\left( {x;{\bf{r}}} \right)  {\bf{n}}\left( \omega  \right)} \right) < 0,\ \forall {\bf r}.
 \end{aligned}
 \end{equation}

 Similar to Section~\ref{subsubsectionSingleMRArea}, for any given placement position $x$ and rotation angle $\omega$, the maximum and minimum projection values within $\cal A$ are respectively defined as
 \begin{equation}\label{areaMaximumDeviationFR}
 {\Delta _{\max }}\left( {x,\omega } \right) \triangleq \mathop {\max }\limits_{{\bf{r}} \in {\cal A}}\ {\left( {{{\bf{a}}_r}\left( {x;{\bf{r}}} \right) - {{\bf{a}}_t}\left( x \right)} \right)^T}{\bf{u}}\left( \omega  \right),
 \end{equation}
 \begin{equation}\label{areaMinimumDeviationFR}
 {\Delta _{\min }}\left( {x,\omega } \right) \triangleq \mathop {\min }\limits_{{\bf{r}} \in {\cal A}}\ {\left( {{{\bf{a}}_r}\left( {x;{\bf{r}}} \right) - {{\bf{a}}_t}\left( x \right)} \right)^T}{\bf{u}}\left( \omega  \right).
 \end{equation}

 In the following, we propose an efficient rotation design. Specifically, for any given placement position $x$, the maximum and minimum projection values without rotation (i.e., $\omega = 0$) can be obtained based on \eqref{areaMaximumDeviation} and \eqref{areaMinimumDeviation}, respectively, with the corresponding area locations denoted as ${{\bf{r}}_{\max }}\left( x \right)$ and ${{\bf{r}}_{\min }}\left( x \right)$, respectively. In order to balance the array factor within $\cal A$, the rotation angle is designed such that $\Delta \left( {x,\omega ;{{\bf{r}}_{\max }}\left( x \right)} \right) =  - \Delta \left( {x,\omega ;{{\bf{r}}_{\min }}\left( x \right)} \right)$, i.e., the boundary points ${{\bf{r}}_{\max }}\left( x \right)$ and ${{\bf{r}}_{\min }}\left( x \right)$ enjoy the identical array factor. Thus, we have
 \begin{equation}
 {\left( {{{\bf{a}}_r}\left( {x;{{\bf{r}}_{\max }}\left( x \right)} \right) + {{\bf{a}}_r}\left( {x;{{\bf{r}}_{\min }}\left( x \right)} \right) - 2{{\bf{a}}_t}\left( x \right)} \right)^T}{\bf{u}}\left( \omega  \right) = 0.
 \end{equation}
 Based on the result in Proposition \ref{rotationAngleproposition}, the rotation angle is given by $\omega \left( x \right) = \arctan \left( { - \frac{{{{\left[ {{{\bf{a}}_r}\left( {x;{{\bf{r}}_{\max }}\left( x \right)} \right) + {{\bf{a}}_r}\left( {x;{{\bf{r}}_{\min }}\left( x \right)} \right) - 2{{\bf{a}}_t}\left( x \right)} \right]}_1}}}{{{{\left[ {{{\bf{a}}_r}\left( {x;{{\bf{r}}_{\max }}\left( x \right)} \right) + {{\bf{a}}_r}\left( {x;{{\bf{r}}_{\min }}\left( x \right)} \right) - 2{{\bf{a}}_t}\left( x \right)} \right]}_2}}}} \right)$. Then, by substituting $\omega \left( {x} \right)$ into \eqref{areaMaximumDeviationFR} and \eqref{areaMinimumDeviationFR}, the maximum and minimum projection values within $\cal A$ for given placement position $x$ and rotation angle $\omega \left( {x} \right)$ can be obtained, denoted as ${\Delta _{\max }}\left( {x,\omega \left( {x} \right)} \right)$ and ${\Delta _{\min}}\left( {x,\omega \left( {x} \right)} \right)$, respectively.

 For electrically large FR, problem \eqref{optimizationProblemSingleFRArea} is reduced to
 \begin{equation}\label{reducedOptimizationProblemSingleFRArea}
 \mathop {\max }\limits_{x }\ \ \mathop {\min }\limits_{{\Delta _{\min }}\left( {x,\omega \left( {x} \right)} \right) \le \Delta  \le {\Delta _{\max }}\left( {x,\omega \left( {x} \right)} \right)} \;\;{f_2}\left( {x;\Delta } \right).\\
 \end{equation}
 The detail for solving \eqref{reducedOptimizationProblemSingleFRArea} is similar to that for solving \eqref{reducedProblemSingleMRArea}, which is omitted for brevity.

 \subsubsection{Multi-FR Case}\label{subsubsectionMultiFRArea}
 For the case of area coverage with electrically large multiple FRs, problem (P1) is reduced to
 \begin{equation}\label{problemAreaMultiFR}
 \begin{aligned}
 &\mathop {\max }\limits_{\left\{ {{x_m},{\omega _m}} \right\}_{m = 1}^M} \mathop {\min }\limits_{{\bf{r}} \in {\cal A}}\ \ \sum\limits_{m = 1}^M {{f_2}\left( {{x_m},{\omega _m};{\bf{r}}} \right)} \\
 {\rm{s.t.}}&\ \left| {{x_m} - {x_n}} \right| \ge {d_{m,n}^{\min }},\ \forall m,n \in {\cal M},\ m \ne n,\\
 &\ \left( {{{\bf{a}}_t^T}\left( {{x_m}} \right)  {\bf{n}}\left( {{\omega _m}} \right)} \right)\left( {{{\bf{a}}_r^T}\left( {{x_m};{\bf{r}}} \right)  {\bf{n}}\left( {{\omega _m}} \right)} \right) < 0,\ \forall m, {\bf r}.
 \end{aligned}
 \end{equation}

 Motivated by the multi-MR design in Section~\ref{subsubsectionMultiMRArea}, we propose a sequential placement and rotation design. Let $x_m^ \star $ and $\omega _m^ \star $ denote the optimized placement position and rotation angle of FR $m$, and ${{\bf r}_{m,l}}$ and ${{\bf r}_{m,r}}$ follow the same definitions as in Section~\ref{subsubsectionMultiMRArea}. The parameters are then determined in the following order,
 \begin{equation}
 {{\bf{r}}_{1,r}} \to x_1^ \star  \to \omega _1^ \star  \to {{\bf{r}}_{1,l}}\left( {{{\bf{r}}_{2,r}}} \right) \to x_2^ \star  \to \omega _2^ \star  \to  \cdots.
 \end{equation}

 Specifically, by letting ${{\bf{r}}_{1,r}} = {{\bf{r}}_{ur}}$, the placement position of the first FR is designed such that the concatenated path loss from the Tx to ${{\bf{r}}_{1,r}}$ is minimized, i.e.,
 \begin{equation}\label{placementFR1}
 x_1^ \star  = \arg \mathop {\min }\limits_x {f_{{\rm{pl}}}}\left( {x;{{\bf{r}}_{1,r}}} \right)\triangleq d_t^2\left( x \right)d_r^2\left( {x;{{\bf{r}}_{1,r}}} \right).
 \end{equation}

 Then, for FR $m$, when the placement position $x_m^ \star$ is obtained, the rotation angle ${{\omega _m}}$ is designed such that
 \begin{equation}\label{rightEndpointEquFR}
 \Delta \left( {x_m^ \star ,{\omega _m};{{\bf{r}}_{m,r}}} \right) = \frac{1}{{2{{\bar L}_1}}}.
 \end{equation}
 The solution to \eqref{rightEndpointEquFR} can be obtained numerically and denoted as ${\omega _m^\star}$. Furthermore, one location corresponding to the left endpoint of the beamforming main lobe w.r.t. FR $m$ can be determined by solving the following equation,
 \begin{equation}\label{leftEndpointEquFR}
 \Delta \left( {x_m^ \star ,\omega _m^ \star ;{{\bf{r}}_{m,l}}} \right) =  - \frac{1}{{2{{\bar L}_1}}}.
 \end{equation}

 Subsequently, the placement position of FR $m+1$ is selected to reduce the concatenated path loss from the Tx to ${{\bf{r}}_{m + 1,r}} = {{\bf{r}}_{m,l}}$, and its rotation angle is designed such that ${\bf{r}}_{m,l}$ is covered by the right endpoint of the beamforming main lobe w.r.t. FR $m+1$. Finally, the sequential placement and rotation design is ended until the lower left corner ${\bf r}_{ll}$ is covered by the beamforming main lobe of one FR. The main procedures are summarized in Algorithm~\ref{alg4}.

 \begin{algorithm}[t]
 \caption{Sequential Placement and Rotation Algorithm for Solving \eqref{problemAreaMultiFR}}
 \label{alg4}
 \begin{algorithmic}[1]
 %\STATE \textbf{Input:} The target area $\cal A$.
 \STATE Initialize ${{\bf{r}}_{1,r}} = {{\bf{r}}_{ur}}$, and let $m=1$.
 \STATE Obtain the optimized placement position of FR 1 based on \eqref{placementFR1}.
 \REPEAT
 \STATE For given ${{\bf r}_{m,r}}$ and $x_m^ \star$, solve the equation \eqref{rightEndpointEquFR}, and denote the solution as $\omega_m^ \star$.
 \STATE For given $x_m^ \star$ and $\omega_m^ \star$, solve the equation \eqref{leftEndpointEquFR}, and denote the solution as ${\bf r}_{m,l}$.
 \STATE ${{\bf{r}}_{m + 1,r}} = {{\bf{r}}_{m,l}}$.
 \STATE \textbf{if} {$m = 1$}
 \STATE \quad $x_{m + 1}^ -  = x_m^ \star  - d\left( {x_m^ \star ,\omega _m^ \star } \right)$.
 \STATE \quad  $x_{m + 1}^ +  = x_m^ \star  + d\left( {x_m^ \star ,\omega _m^ \star } \right)$.
 \STATE \textbf{else}
 \STATE \quad \textbf{if} $x_m^ \star  = x_m^ - $
 \STATE \quad \quad $x_{m + 1}^ -  = x_m^ \star  - d\left( {x_m^ \star ,\omega _m^ \star } \right)$.
 \STATE \quad \quad $x_{m + 1}^ +  = x_m^ + $.
 \STATE \quad \textbf{else} %$x_m^ \star  = x_m^ + $
 \STATE \quad \quad $x_{m + 1}^ -  = x_m^ - $.
 \STATE \quad \quad $x_{m + 1}^ +  = x_m^ \star  + d\left( {x_m^ \star ,\omega _m^ \star } \right)$.
 \STATE \quad \textbf{end if}
 \STATE \textbf{end if}
 \STATE \textbf{if} ${f_{{\rm{pl}}}}\left( {x_{m + 1}^ - ;{{\bf{r}}_{m + 1,r}}} \right)\le {f_{{\rm{pl}}}}\left( {x_{m + 1}^ + ;{{\bf{r}}_{m + 1,r}}} \right)$
 \STATE \quad $x_{m + 1}^ \star  = x_{m + 1}^ - $.
 \STATE \textbf{else}
 \STATE \quad $x_{m + 1}^ \star  = x_{m + 1}^ + $.
 \STATE \textbf{end if}
 \STATE Update $m = m+1$.
 \UNTIL ${\bf r}_{ll}$ is covered by the beamforming main lobe of one FR.
 \end{algorithmic}
 \end{algorithm}

%% >>>>>>>>>>>>>SECTIONS V -  here >>>>>>>>>>>>
 \section{Numerical Results}\label{sectionNumericalResults}
 In this section, numerical results are provided to evaluate the proposed MR and FR designs. The carrier frequency is ${f_c} = 2.4$ GHz. The length and width of the reflector are ${L_1} = {{\bar L}_1}\lambda  = 10\lambda  = 1.25$ m and ${L_2} = {{\bar L}_2}\lambda  = 5\lambda  = 0.625$ m, respectively. Unless otherwise stated, the transmit power is ${P_t} = 30$ dBm.

 \subsection{Single Target Location}
 \begin{figure}[!t]
 \centering
 \centerline{\includegraphics[width=3.2in,height=2.4in]{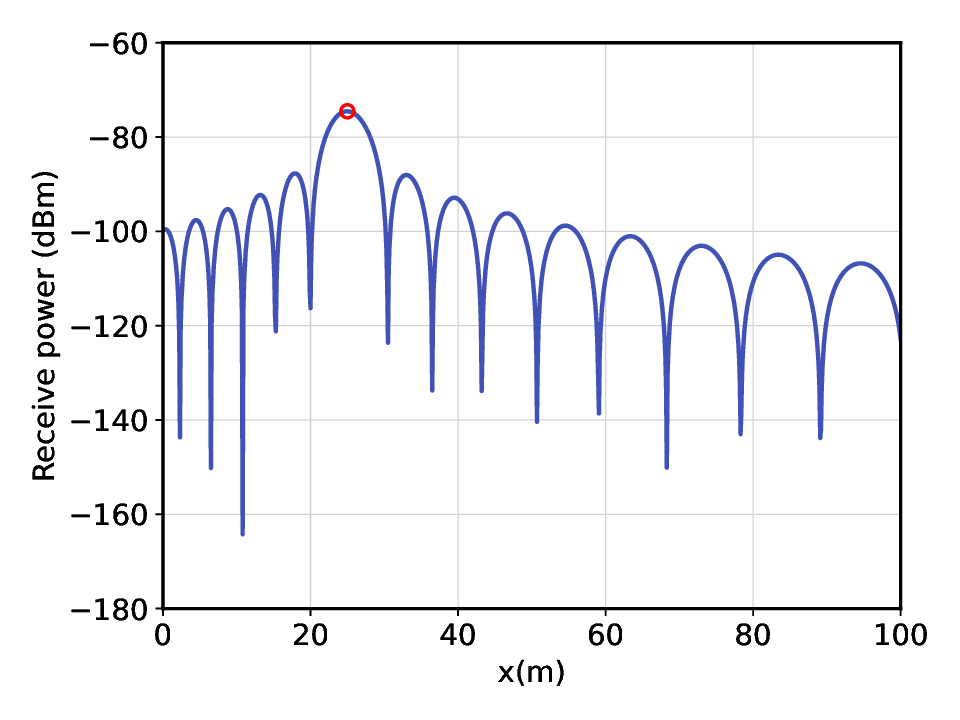}}
 \caption{The receive power versus the placement position $x$ for single-MR aided single target location power enhancement, where the placement position given in \eqref{maximumReceivedPowerLocation} is marked by the red circle.}
 \label{fig:receivedPowerVersusPlacementSP}
 \end{figure}

 Fig.~\ref{fig:receivedPowerVersusPlacementSP} shows the receive power of the Rx versus the placement position $x$ for MR, where the locations of the Tx and Rx are ${\bf{t}} = {\left[ {0, - 50} \right]^T}$ m and ${\bf{r}} = {\left[ {100, - 150} \right]^T}$ m, respectively. It is observed that the receive power exhibits significant variations as the placement position changes, and the performance gap between the maximum and minimum receive power is up to tens of dBm. This demonstrates the necessity of placement optimization for MR. It is also observed that when ${{\bar L}_1} \gg 1$, the derived placement position in \eqref{maximumReceivedPowerLocation} gives a quite accurate approximation of the optimal placement position to maximize the receive power at the Rx.

 \begin{figure}[!t]
 \centering
 \centerline{\includegraphics[width=3.2in,height=2.4in]{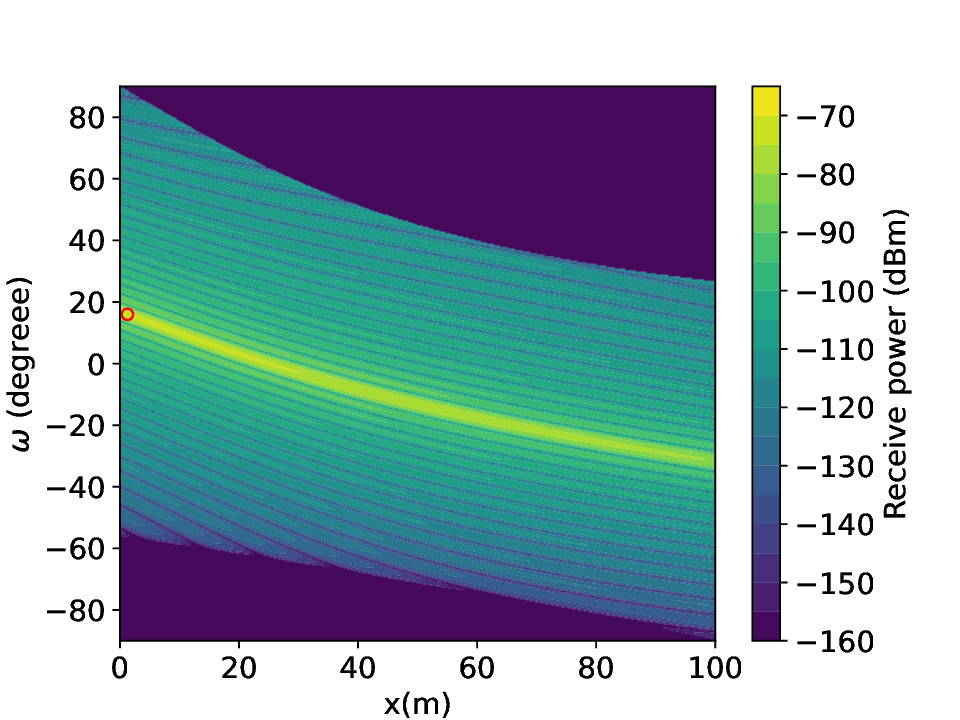}}
 \caption{The receive power versus the placement position $x$ and rotation angle $\omega$  for single-FR aided single target location power enhancement, where the optimal placement position and rotation angle pair $\left( {x^{\star},\omega^{\star} } \right)$ is marked by the red circle.}
 \label{fig:receivedPowerVersusPlacementRotationSP}
 %\vspace{-0.5cm}
 \end{figure}

 Fig.~\ref{fig:receivedPowerVersusPlacementRotationSP} shows the receive power versus the placement position $x$ and rotation angle $\omega$ for FR, where the receive power below $-160$ dBm is truncated to $-160$ dBm for convenience of presentation. It is observed that a significant performance gain can be achieved by adjusting the placement position and rotation angle. Moreover, for the considered setup, different from the MR that is placed at the specular reflection point, the FR is preferable to deploy in the vicinity of $x$-coordinate of the Tx.

\begin{figure}[!t]
 \centering
 \centerline{\includegraphics[width=3.2in,height=2.4in]{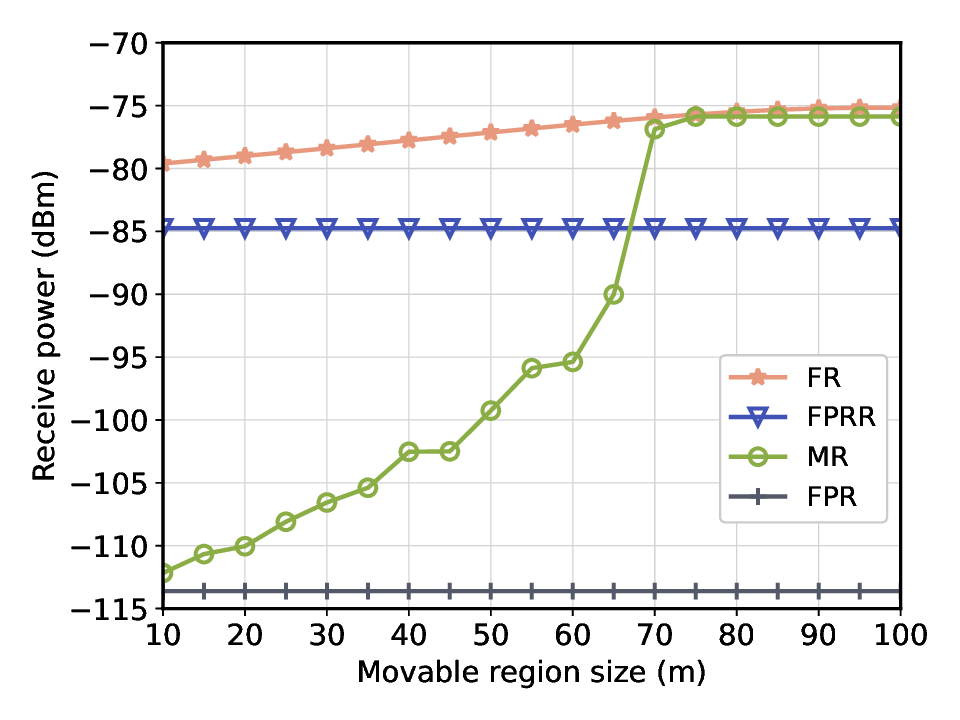}}
 \caption{The receive power versus the movable region size.}
 \label{fig:receivedPowerVersusRegionSizeSP}
 \end{figure}

 Fig.~\ref{fig:receivedPowerVersusRegionSizeSP} shows the receive power versus the movable region size. For comparison, the following two benchmark schemes are considered: 1) Fixed-position reflector (FPR): the reflector placement position is fixed at $x = {x_t}$; 2) Fixed-position rotatable reflector (FPRR): the reflector is capable of adjusting the rotation angle, while its placement position is fixed at $x = {x_t}$. The locations of the Tx and Rx are ${\bf{t}} = {\left[ {0, - 150} \right]^T}$ m and ${\bf{r}} = {\left[ {100, - 60} \right]^T}$ m, respectively. It is observed that the FR achieves the best performance, and the receive power for both the FR and MR improves as the movable region size increases. This is expected since the MR is of high probability to move to the specular point for a larger movable region size, and the FR is more likely to be close to the optimal placement position. By contrast, the receive power for both the FPR and FPRR remains unchanged as the region size increases, which is due to the fact that the FPR and FPRR cannot exploit the spatial design DoF of placement. Moreover, the FPRR yields a much higher receive power than FPR, thanks to the additional rotation angle adjustment. In particular, considerable performance gain is achieved by FPRR over the MR for a relatively small movable region size, which may provide a satisfied scheme for the case of the restricted movable region.

 \begin{figure}[!t]
 \centering
 \centerline{\includegraphics[width=3.2in,height=2.4in]{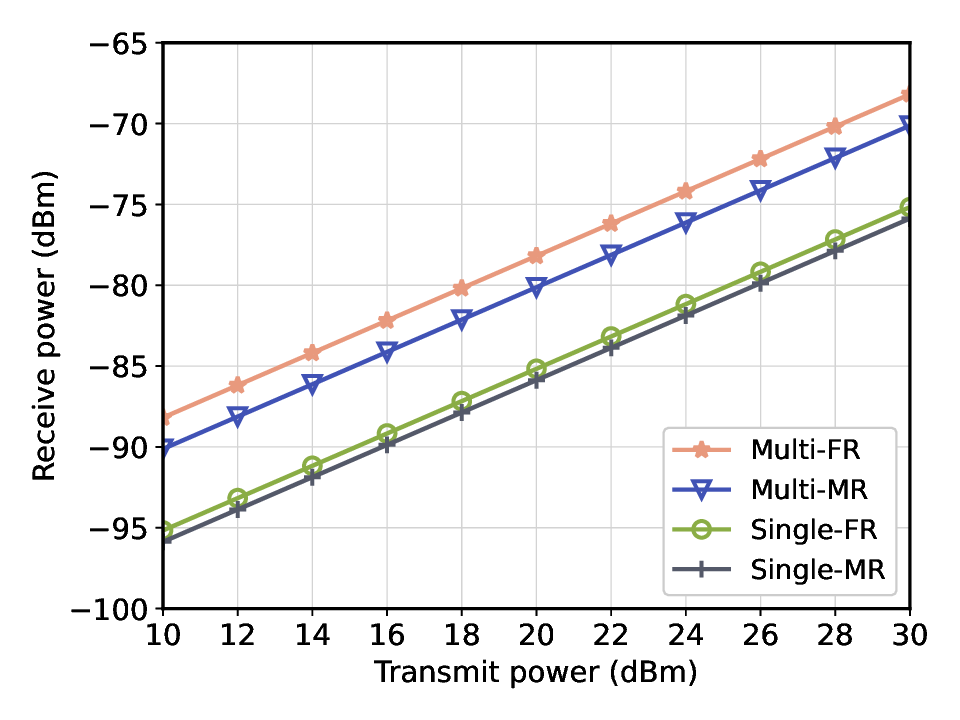}}
 \caption{The receive power versus the transmit power for multi-MR and multi-FR aided single target location power enhancement.}
 \label{fig:receivedPowerVersusTransmitPowerSPMulti}
 \end{figure}

 As a further comparison, Fig.~\ref{fig:receivedPowerVersusTransmitPowerSPMulti} shows the receive power versus the transmit power. For the cases of multi-MR and multi-FR, the number of reflectors is $M =5$. It is observed that multi-FR achieves a higher receive power than single-FR, and a similar result can be observed for the case of MR. This is due to the fact that more power is captured and reflected by multiple reflectors. It is also observed that compared to the case of the single reflector, the performance gain of FR over MR is more significant in the case of multiple reflectors. This is expected since the multi-FR is able to obtain the full array factor at multiple placement positions, by flexibly adjusting the rotation angle.

 \subsection{Area Coverage Enhancement}
 \begin{figure}[!t]
 \centering
 \centerline{\includegraphics[width=3.2in,height=2.4in]{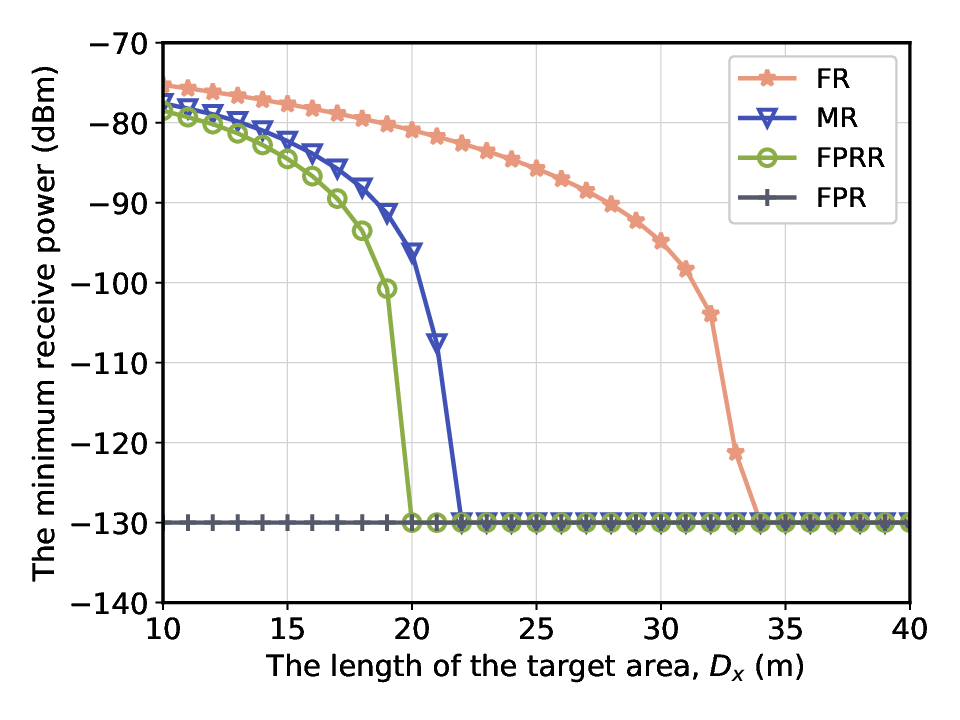}}
 \caption{The minimum receive power versus the length of the target area.}
 \label{fig:minimumReceivedPowerVersusAreaLength}
 \end{figure}

 Fig.~\ref{fig:minimumReceivedPowerVersusAreaLength} shows the minimum receive power versus the length of the target area $D_x$ by considering the single reflector, where the receive power below $-130$ dBm is truncated to $-130$ dBm for convenience of presentation. The center of the target area is ${{\bf{r}}_c} = {\left[ {100, - 75} \right]^T}$ m, and its width is fixed as ${D_y} = 10$ m. The location of the Tx is ${\bf t} = {\left[ {0, - 50} \right]^T}$ m. The benchmark schemes of FPR and FPRR are considered, where the reflector is placed above the midpoint between the Tx and area center, i.e., $x = \left( {{x_t} + {x_c}} \right)/2$. It is observed that the minimum receive power of the FR outperforms those of other three schemes for a relatively small area length, thanks to its flexible placement and rotation adjustment. Besides, the minimum receive power of FPR is always equal to zero (i.e., $-130$ dBm), this is because its null-to-null beam cannot cover the whole target area. It is also observed that as $D_x$ increases, the minimum receive power of FPRR, MR and FR decreases and eventually drops to zero successively. The above result indicates that the FR is able to increase the coverage area size as compared to other three schemes, by fully exploiting the spatial design DoFs of placement and rotation.

 \begin{figure}
 \centering
 \subfigure[Proposed scheme]{
 \begin{minipage}[t]{0.5\textwidth}
 \centering
 \centerline{\includegraphics[width=3.2in,height=2.4in]{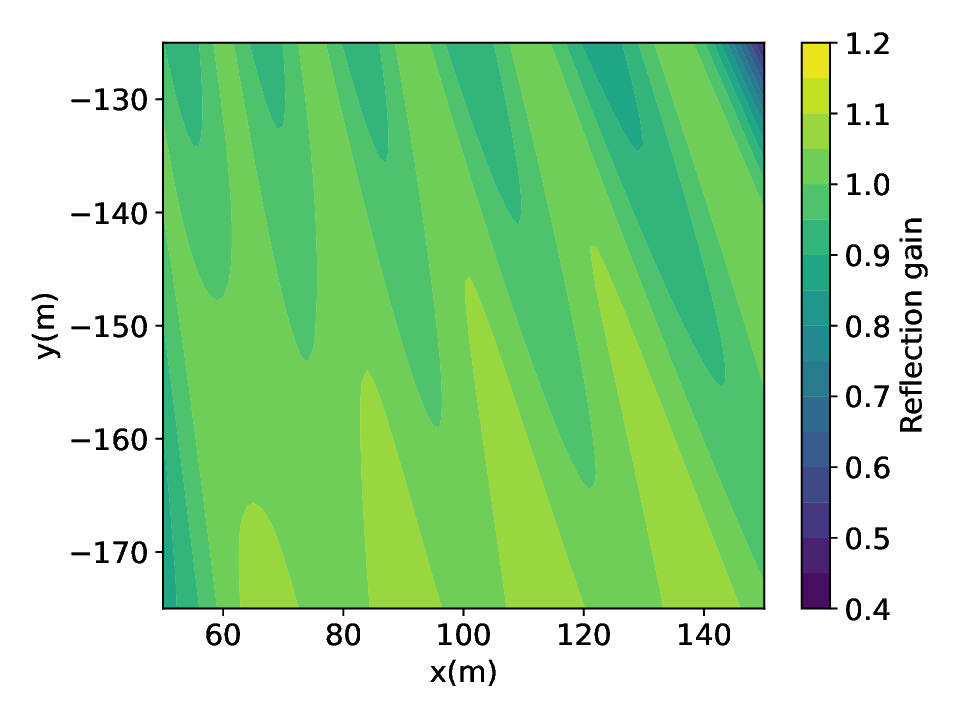}}
 \end{minipage}
 }
 \subfigure[Benchmark scheme]{
 \begin{minipage}[t]{0.5\textwidth}
 \centering
 \centerline{\includegraphics[width=3.2in,height=2.4in]{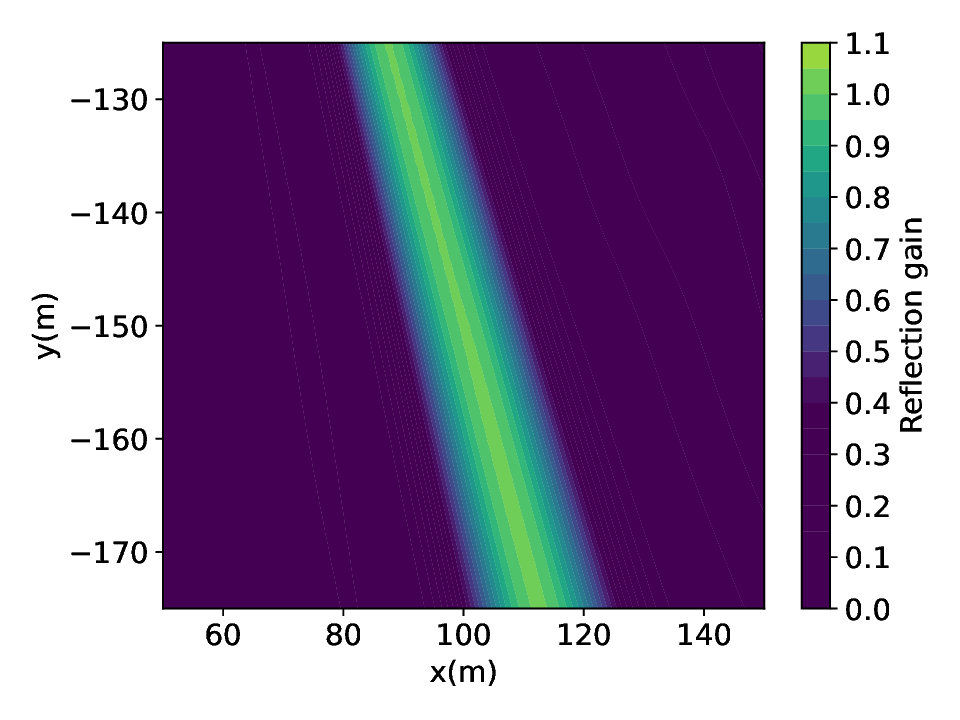}}
 \end{minipage}
 }
 \caption{Reflection gain at different Rx locations for multi-MR aided area coverage enhancement.}
 \label{fig:reflectionGainVersusLocation}
 \vspace{-0.3cm}
 \end{figure}

 Fig.~\ref{fig:reflectionGainVersusLocation} shows the reflection gain at different Rx locations for multi-MR aided area coverage enhancement, where the reflection gain at location $\bf r$ is defined as $\sum\nolimits_m {{f_2}\left( {{x_m};{\bf{r}}} \right)} $. The center of the target area is ${{\bf{r}}_c} = {\left[ {100, - 150} \right]^T}$ m, and its length and width are ${D_x} = 100$ m and ${D_y} = 50$ m, respectively. For comparison, the benchmark scheme where the reflectors are equally placed within the range $\left[ {{x_t},{x_c} + {D_x}/2} \right]$ is considered. Based on Algorithm~\ref{alg2}, the number of required reflector is $M = 7$. It is observed from Fig.~\ref{fig:reflectionGainVersusLocation}(a) that for the proposed sequential placement scheme in Algorithm~\ref{alg2}, any location of the target area enjoys a reflection gain larger than $0.4$, which is value of the endpoint of the beamforming main lobe. This implies that with the proposed sequential placement algorithm, any location of the target area can be covered by the beamforming main lobe of at least one reflector. By contrast, for the benchmark scheme, there exist many coverage holes with zero reflection gain, as can be seen in Fig.~\ref{fig:reflectionGainVersusLocation}(b).

 \begin{figure}[!t]
 \centering
 \centerline{\includegraphics[width=3.2in,height=2.4in]{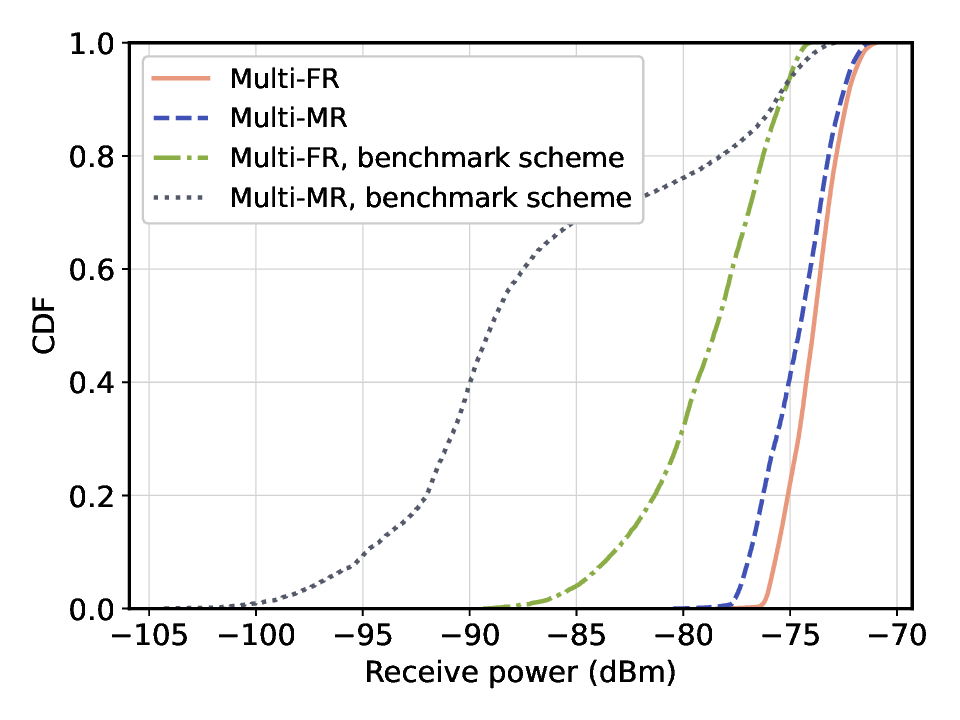}}
 \caption{The CDF of the receive power within the target area.}
 \label{fig:CDFofReceivedPower}
 \vspace{-0.3cm}
 \end{figure}

 Fig.~\ref{fig:CDFofReceivedPower} shows the  cumulative distribution function (CDF) of the receive power within the target area for the case of multiple reflectors. The number of MRs and FRs required for coverage are $7$ and $6$, respectively. As a comparison, the benchmark scheme of multi-MR in Fig.~\ref{fig:reflectionGainVersusLocation} is considered. For the benchmark scheme of multi-FR, the FRs are equally placed within $\left[ {{x_t},{x_c} + {D_x}/2} \right]$, and the rotation angles are designed similar to Algorithm~\ref{alg4}. It is observed that the proposed multi-FR and multi-MR schemes are superior to their counterparts in the benchmark schemes, as expected. Moreover, FR in general yields a better area coverage performance compared to MR, even with a smaller number of reflectors, thanks to the additional design DoF of rotation.

%% >>>>>>>>>>>>>SECTIONS VI -  here >>>>>>>>>>>>
\section{Conclusion}\label{sectionConclusion}
 This paper proposed a new paradigm of passive reflective beamforming enabled by FR, which enables a flexible beamforming direction via both placement position and rotation angle adjustments. The minimum expected receive power over all locations within a given target area was maximized by jointly optimizing FRs' placement positions and rotation angles. To gain useful insights, the special case of MR was first considered by fixing its rotation angle, where the single- and multi-MR aided coverage enhancement were respectively studied. Moreover, the MR design was extended to the general case of the single- and multi-FR aided coverage enhancement. Numerical results showed that the proposed FR scheme significantly outperforms benchmark schemes in terms of receive power enhancement. Despite the promising advantages and performance enhancement, there exist several challenges for wireless communication with FR. For example, since joint placement position and rotation angle optimization relies on channel state information (CSI), the efficient CSI acquisition method and the impact of CSI errors on system performance are worthwhile to investigate in the future. Besides, inter-user interference (IUI) mitigation for multi-user communication assisted by FR, and extensions of this work to the more general multi-path channels and MIMO communications deserve further studies.

\begin{appendices}
\section{Proof of Proposition \ref{maximumReceivedPowerProposition}}\label{proofmaximumReceivedPowerProposition}
 When $L_1$ is much larger than the signal wavelength, the sinc function changes rapidly, as shown in Fig.~\ref{fig:fFunctionVersusPlacement}(a). To this end, the optimization problem \eqref{optimizationProblemSPSingle} can be approximated to maximize the array factor $f_2\left(x\right)$, yielding the following problem
 \begin{equation}\label{sincMaximizationProblemSTL}
 \begin{aligned}
 \mathop {{\rm{max}}}\limits_{x} &\ \  {\rm{sin}}{{\rm{c}}^2}\left( {\pi {{\bar L}_1}\left( {\frac{{{x_r} - x}}{{\left\| {{\bf{r}} - {\bf{q}}} \right\|}} - \frac{{x - {x_t}}}{{\left\| {{\bf{q}} - {\bf{t}}} \right\|}}} \right)} \right).
 \end{aligned}
 \end{equation}

 It is known that the function ${\rm sinc}\left(x\right)$ achieves the maximum value at $x = 0$. Thus, problem \eqref{sincMaximizationProblemSTL} can be maximized by letting $\frac{{{x_r} - x}}{{\left\| {{\bf{r}} - {\bf{q}}} \right\|}} - \frac{{x - {x_t}}}{{\left\| {{\bf{q}} - {\bf{t}}} \right\|}} = 0$. With ${\bf{q}} = {\left[ {x,0,0} \right]^T}$, ${\bf{t}} = {\left[ {{x_t},{y_t},0} \right]^T}$ and ${\bf{r}} = {\left[ {{x_r},{y_r},0} \right]^T}$, we have
 \begin{equation}
 \frac{{{x_r} - x}}{{\sqrt {{{\left( {{x_r} - x} \right)}^2} + y_r^2} }} - \frac{{x - {x_t}}}{{\sqrt {{{\left( {x - {x_t}} \right)}^2} + y_t^2} }} = 0.
 \end{equation}
 It can be seen that the solution $x$ is within $\left[ {{x_t},{x_r}} \right]$. After some manipulations, we have
 \begin{equation}
  \frac{{{y_r}}}{{{x_r} - x}} = \frac{{{y_t}}}{{x - {x_t}}},
 \end{equation}
 and the optimal placement is
 \begin{equation}
 x^{\star} = {x_t} + \frac{{{y_t}}}{{{y_t} + {y_r}}}\left( {{x_r} - {x_t}} \right).
 \end{equation}
 This thus completes the proof of Proposition \ref{maximumReceivedPowerProposition}.

\section{Proof of Lemma \ref{maximumReceivedPowerSmalllemma}}\label{proofOfmaximumReceivedPowerSmalllemma}
 Since the objective function of \eqref{sincMaximizationProblemSmallSTL} is differentiable, the extreme points can be obtained by letting its first-order derivative w.r.t. $x$ equal to zero, which yields
 \begin{equation}\label{firstOrderDerivative}
 2\left( {{{\left( {{x_r} - x} \right)}^2} + y_r^2} \right)\left( {{a_c}{x^3} + {b_c}{x^2} + {c_c}x + {d_c}} \right) = 0,
 \end{equation}
 where $a_c$, $b_c$, $c_c$, and $d_c$ are given below \eqref{smallDerivationEquation}. It is observed that ${{{\left( {{x_r} - x} \right)}^2} + y_r^2} > 0$, and thus \eqref{firstOrderDerivative} is equivalent to \eqref{smallDerivationEquation}.
 The proof of Lemma \ref{maximumReceivedPowerSmalllemma} is thus completed.

\section{Proof of Lemma \ref{leftEndPointlemma}}\label{proofOfleftEndPointlemma}
 Depending on the relationship between ${\Delta}\left( {x_m^ \star ;{{\bf{r}}_{ul}}} \right)$ and ${ - \frac{1}{{2{{\bar L}_1}}}}$, we have the following two cases.

 \begin{figure}[!t]
 \centering
 \subfigure[${\Delta}\left( {x_m^ \star ;{{\bf{r}}_{ul}}} \right) <  - \frac{1}{{2{{\bar L}_1}}}$]{
 \begin{minipage}[t]{0.5\textwidth}
 \centering
 \centerline{\includegraphics[width=2.5in,height=1.8in]{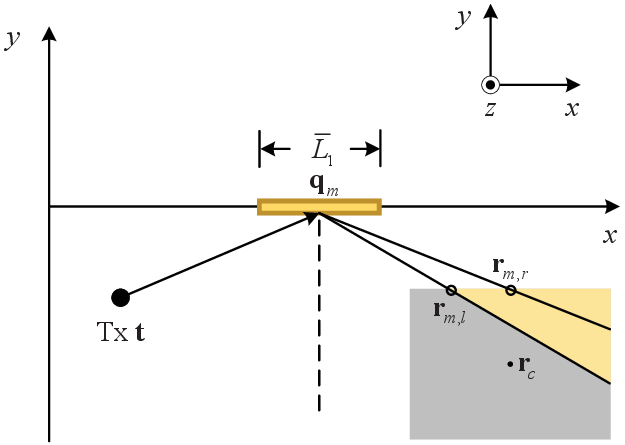}}
 \end{minipage}
 }
 \subfigure[${\Delta }\left( {x_m^ \star ;{{\bf{r}}_{ul}}} \right) \ge  - \frac{1}{{2{{\bar L}_1}}}$ and ${\Delta}\left( {x_m^ \star ;{{\bf{r}}_{ll}}} \right) <  - \frac{1}{{2{{\bar L}_1}}}$]{
 \begin{minipage}[t]{0.5\textwidth}
 \centering
 \centerline{\includegraphics[width=2.5in,height=1.8in]{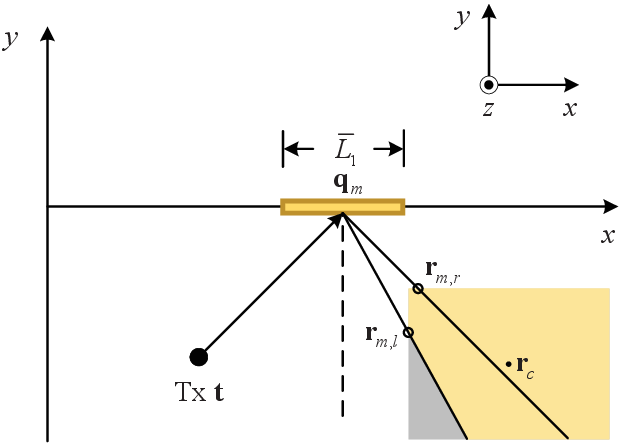}}
 \end{minipage}
 }
 \caption{An illustration of the coverage area, where the yellow part of the target area has been covered by the beamforming main lobes of the first $m$ MRs, and ${{\bf{r}}_{m,l}}$ and ${{\bf{r}}_{m,r}}$ denote the locations corresponding to left and right endpoints of the beamforming main lobe w.r.t. MR $m$, respectively.}
 \label{illustrationOFCoverageArea}
 \end{figure}

 \emph{Case 1:} ${\Delta}\left( {x_m^ \star ;{{\bf{r}}_{ul}}} \right) <  - \frac{1}{{2{{\bar L}_1}}}$, i.e., the upper left corner ${\bf r}_{ul}$ is not covered by the beamforming main lobe of MR $m$, as illustrated in Fig.~\ref{illustrationOFCoverageArea}(a). %This implies that the $x$-coordinate value of the location corresponding to the left endpoint of the beamforming main lobe is larger than that of ${\bf r}_{\rm ul}$, and
 Then, a location corresponding to the left endpoint of the beamforming main lobe is chosen as ${\left[ {{x_r},{y_c} + {D_y}/2} \right]^T}$, where ${x_c} - {D_x}/2 < {x_r} < {x_c} + {D_x}/2$.%, as illustrated in Fig.~\ref{illustrationOFCoverageArea}(a).

 \emph{Case 2:} ${\Delta}\left( {x_m^ \star ;{{\bf{r}}_{ul}}} \right) \ge  - \frac{1}{{2{{\bar L}_1}}}$ and ${\Delta}\left( {x_m^ \star ;{{\bf{r}}_{ll}}} \right) <  - \frac{1}{{2{{\bar L}_1}}}$, i.e., the upper left corner ${\bf r}_{ ul}$ is covered by the beamforming main lobe of MR $m$, as illustrated in Fig.~\ref{illustrationOFCoverageArea}(b). In this case, a location corresponding to the left endpoint of the beamforming main lobe is chosen as ${\left[ {{x_c} - {D_x}/2,{y_r}} \right]^T}$, where ${y_c} - {D_y}/2 < {y_r} < {y_c} + {D_y}/2$.

 By combining the above two cases, the proof of Lemma \ref{leftEndPointlemma} is thus completed.

\section{Proof of Proposition \ref{rotationAngleproposition}}\label{proofOfrotationAngleproposition}
 Since the objective function of problem \eqref{optimizationProblemSingleLocationSFRS1} is maximized when $\Delta \left( {x,\omega } \right) = 0$, we show that for any given placement position $x$, there always exists a rotation angle satisfying the constraint of \eqref{optimizationProblemSingleLocationSFRS1} such that $\Delta \left( {x,\omega } \right) = 0$. Specifically, to ensure that $\Delta \left( {x,\omega } \right) = 0$, we have
 \begin{equation}\label{zeroDeviationFR}
 {\left[ {{{\bf{a}}_r}\left( {x } \right) - {{\bf{a}}_t}\left( x \right)} \right]_1}\cos \omega  + {\left[ {{{\bf{a}}_r}\left( {x} \right) - {{\bf{a}}_t}\left( x \right)} \right]_2}\sin \omega  = 0,
 \end{equation}
 where ${\left[ {{{\bf{a}}_r}\left( {x} \right) - {{\bf{a}}_t}\left( x \right)} \right]_i}$ denotes the $i$-th element of the deflection vector $\left({{{\bf{a}}_r}\left( {x} \right) - {{\bf{a}}_t}\left( x \right)}\right)$. A solution to \eqref{zeroDeviationFR} can be obtained by letting
 \begin{equation}
 \left\{ {\begin{split}
 &{\cos \omega  =  - \frac{{{{\left[ {{{\bf{a}}_r}\left( {x} \right) - {{\bf{a}}_t}\left( x \right)} \right]}_2}}}{{\left\| {{{\bf{a}}_r}\left( {x } \right) - {{\bf{a}}_t}\left( x \right)} \right\|}},}\\
 &{\sin \omega  = \frac{{{{\left[ {{{\bf{a}}_r}\left( {x} \right) - {{\bf{a}}_t}\left( x \right)} \right]}_1}}}{{\left\| {{{\bf{a}}_r}\left( {x} \right) - {{\bf{a}}_t}\left( x \right)} \right\|}}.}
 \end{split}} \right.
 \end{equation}
 The rotation angle is then given by
 \begin{equation}\label{zeroDeviationRotationAngleProof}
 \omega  = \arctan \left( { - \frac{{{{\left[ {{{\bf{a}}_r}\left( {x} \right) - {{\bf{a}}_t}\left( x \right)} \right]}_1}}}{{{{\left[ {{{\bf{a}}_r}\left( {x} \right) - {{\bf{a}}_t}\left( x \right)} \right]}_2}}}} \right),
 \end{equation}
 i.e., \eqref{zeroDeviationRotationAngle}.

 Moreover, with \eqref{zeroDeviationRotationAngleProof}, the normal direction vector is
 \begin{equation}\label{normalDirectionVector}
 {\bf{n}}\left( \omega  \right) =  - \frac{{{{\bf{a}}_r}\left( {x} \right) - {{\bf{a}}_t}\left( x \right)}}{{\left\| {{{\bf{a}}_r}\left( {x} \right) - {{\bf{a}}_t}\left( x \right)} \right\|}}.
 \end{equation}
 Thus, we have
 \begin{equation}
 \begin{aligned}
 {\bf{a}}_t^T\left( x \right){\bf{n}}\left( \omega  \right) & =  - \frac{{\left( {{\bf{a}}_t^T\left( x \right){{\bf{a}}_r}\left( x \right) - {\bf{a}}_t^T\left( x \right){{\bf{a}}_t}\left( x \right)} \right)}}{{\left\| {{{\bf{a}}_r}\left( x \right) - {{\bf{a}}_t}\left( x \right)} \right\|}}\\
 &=  - \frac{{\left( {{\bf{a}}_t^T\left( x \right){{\bf{a}}_r}\left( x \right) - 1} \right)}}{{\left\| {{{\bf{a}}_r}\left( x \right) - {{\bf{a}}_t}\left( x \right)} \right\|}}.
 \end{aligned}
 \end{equation}
 Since ${\bf{a}}_t^T\left( x \right){{\bf{a}}_r}\left( x \right) < 1$, we have ${{\bf{a}}_t^T}\left( x \right) {\bf{n}}\left( \omega  \right) >0$. Similarly, ${{\bf{a}}_r^T}\left( {x} \right) {\bf{n}}\left( \omega  \right) < 0$. Then, the rotation angle given in \eqref{zeroDeviationRotationAngleProof} achieves $\Delta \left( {x,\omega } \right) = 0$, while satisfying the constraint of \eqref{optimizationProblemSingleLocationSFRS1}. The proof of Proposition \ref{rotationAngleproposition} is thus completed.

 \section{Proof of Lemma \ref{distanceAntiBlockageFRlemma}}\label{proofOfDistanceAntiBlockageFR}

 \begin{figure}[!t]
 \centering
 \subfigure[The minimum distance ensuring that the signal reflected by any location\newline of the FR can reach the Rx]{
 \begin{minipage}[t]{0.5\textwidth}
 \centering
 \centerline{\includegraphics[width=2.75in,height=1.8in]{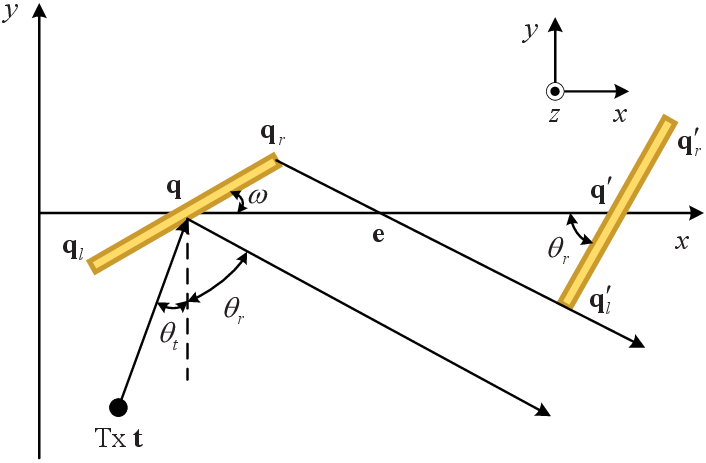}}
 \end{minipage}
 }
 \subfigure[The minimum distance ensuring that the signal from the Tx can reach\newline any location of the FR]{
 \begin{minipage}[t]{0.5\textwidth}
 \centering
 \centerline{\includegraphics[width=2.75in,height=1.8in]{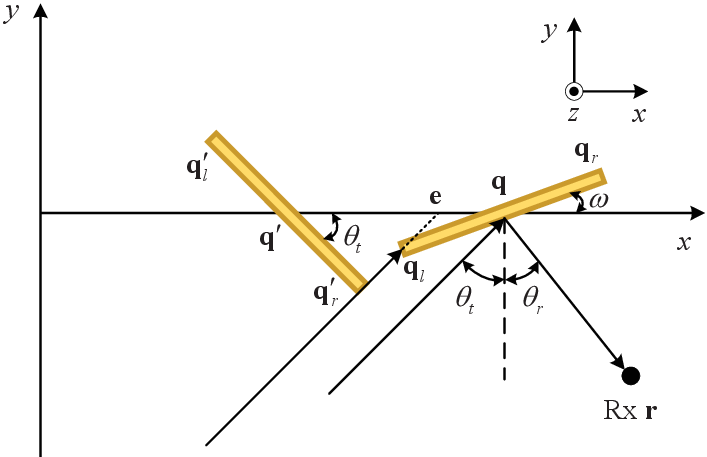}}
 \end{minipage}
 }
 \caption{An illustration of the minimum distance to avoid the signal blockage between adjacent FRs.}
 \label{fig:distanceAntiBlockageFR}
 \end{figure}

 To avoid the signal blockage between adjacent FRs, the inter-FR distance should ensure that the signal from the Tx can reach any location of the FR, and the signal reflected by any location of the FR can reach the Rx. Specifically, we first derive the minimum distance ensuring that the signal reflected by any location of the FR can reach the Rx. As illustrated in Fig.~\ref{fig:distanceAntiBlockageFR}(a), for given FR with the placement position and rotation angle pair being $\left( {x,\omega } \right)$, the location of the right endpoint is denoted as ${{\bf{q}}_r}$.
 %${{\bf{q}}_r} = {\bf{q}} + {\left[ {\frac{{{L_1}}}{2}\cos \omega ,\frac{{{L_1}}}{2}\sin \omega } \right]^T}$. 
 For the standard uniform plane wave (UPW) model, let $\bf e$ denote the intersection point between the $x$-axis and the signal arriving at the Rx via ${{\bf{q}}_r}$. Thus, to avoid the signal blockage, any location of another FR should locate above the straight line formed by ${{\bf{q}}_r}$ and $\bf e$. In particular, the minimum distance satisfying the above condition occurs when ${{\bf{q}}'_r} - {{\bf{q}}'_l}$ is perpendicular to ${{\bf{q}}'_l} - {{\bf{q}}_r}$, as illustrated in Fig.~\ref{fig:distanceAntiBlockageFR}(a). Thus, based on the geometric relationship, the minimum distance is given by ${d_r}\left( {x, \omega}\right) = \left\| {{\bf{e}} - {\bf{q}}} \right\| + \left\| {{\bf{q'}} - {\bf{e}}} \right\| = \frac{{{L_1}}}{2}\left( {\cos \omega  + \tan {\theta _r}\sin \omega  + \frac{1}{{\cos {\theta _r}}}} \right)$.

 On the other hand, as illustrated in Fig.~\ref{fig:distanceAntiBlockageFR}(b), the minimum distance ensuring that the signal from the Tx can reach any location of the FR is similarly obtained, given by ${d_i}\left( {x,\omega } \right) = \frac{{{L_1}}}{2}\left( {\cos \omega  - \tan {\theta _t}\sin \omega  + \frac{1}{{\cos {\theta _t}}}} \right)$. The details are omitted for brevity. Thus, to minimum distance to avoid the signal blockage is $d\left( {x,\omega } \right) = \max \left( {{d_i}\left( {x,\omega } \right),{d_r}\left( {x,\omega } \right)} \right)$. This thus completes the proof of Lemma \ref{distanceAntiBlockageFRlemma}.

\end{appendices}

%\ifCLASSOPTIONcaptionsoff
%  \newpage
%\fi

\bibliographystyle{IEEEtran}
\bibliography{refMovableReflector}

\begin{thebibliography}{10}
\providecommand{\url}[1]{#1}
\csname url@samestyle\endcsname
\providecommand{\newblock}{\relax}
\providecommand{\bibinfo}[2]{#2}
\providecommand{\BIBentrySTDinterwordspacing}{\spaceskip=0pt\relax}
\providecommand{\BIBentryALTinterwordstretchfactor}{4}
\providecommand{\BIBentryALTinterwordspacing}{\spaceskip=\fontdimen2\font plus
\BIBentryALTinterwordstretchfactor\fontdimen3\font minus
  \fontdimen4\font\relax}
\providecommand{\BIBforeignlanguage}[2]{{%
\expandafter\ifx\csname l@#1\endcsname\relax
\typeout{** WARNING: IEEEtran.bst: No hyphenation pattern has been}%
\typeout{** loaded for the language `#1'. Using the pattern for}%
\typeout{** the default language instead.}%
\else
\language=\csname l@#1\endcsname
\fi
#2}}
\providecommand{\BIBdecl}{\relax}
\BIBdecl

\bibitem{lu2024tutorial}
H.~Lu \emph{et~al.}, ``{A tutorial on near-field XL-MIMO communications towards
  6G},'' \emph{IEEE Commun. Surv. Tuts.}, vol.~26, no.~4, pp. 2213--2257, 4th
  Quart. 2024.

\bibitem{bjornson2019massive}
E.~Bj{\"o}rnson, L.~Sanguinetti, H.~Wymeersch, J.~Hoydis, and T.~L. Marzetta,
  ``Massive {MIMO} is a reality -- what is next?: Five promising research
  directions for antenna arrays,'' \emph{Digit. Signal Process.}, vol.~94, pp.
  3--20, Nov. 2019.

\bibitem{wang2024extremely}
Z.~Wang, J.~Zhang, H.~Du, W.~E.~I. Sha, B.~Ai, D.~Niyato, and M.~Debbah,
  ``Extremely large-scale {MIMO}: Fundamentals, challenges, solutions, and
  future directions,'' \emph{IEEE Wireless Commun.}, vol.~31, no.~3, pp.
  117--124, Jun. 2024.

\bibitem{ITU}
``Framework and overall objectives of the future development of {IMT} for 2030
  and beyond,'' ITU-R, DRAFT NEW RECOMMENDATION, Jun. 2023.

\bibitem{Han2023Towards}
Y.~Han, S.~Jin, M.~Matthaiou, T.~Q.~S. Quek, and C.-K. Wen, ``{Towards extra
  large-scale MIMO: New channel properties and low-cost designs},'' \emph{IEEE
  Internet Things J.}, vol.~10, no.~16, pp. 14\,569--14\,594, Aug. 2023.

\bibitem{wang2023can}
H.~Wang and Y.~Zeng, ``Can sparse arrays outperform collocated arrays for
  future wireless communications?'' in \emph{Proc. IEEE GLOBECOM Workshops (GC
  Wkshps)}, 2023, pp. 667--672.

\bibitem{wang2024enhancing}
H.~Wang, C.~Feng, Y.~Zeng, S.~Jin, C.~Yuen, B.~Clerckx, and R.~Zhang,
  ``Enhancing spatial multiplexing and interference suppression for near-and
  far-field communications with sparse {MIMO},'' \emph{arXiv preprint
  arXiv:2408.01956}, 2024.

\bibitem{lu2024group}
H.~Lu, Y.~Zeng, S.~Jin, and R.~Zhang, ``Group movable antenna with flexible
  sparsity: Joint array position and sparsity optimization,'' \emph{IEEE
  Wireless Commun. Lett.}, vol.~13, no.~12, pp. 3573--3577, Dec. 2024.

\bibitem{li2024sparse}
X.~Li, H.~Min, Y.~Zeng, S.~Jin, L.~Dai, Y.~Yuan, and R.~Zhang, ``{Sparse MIMO
  for ISAC: New opportunities and challenges},'' \emph{IEEE Wireless Commun.},
  2024.

\bibitem{pal2010nested}
P.~Pal and P.~P. Vaidyanathan, ``Nested arrays: A novel approach to array
  processing with enhanced degrees of freedom,'' \emph{IEEE Trans. Signal
  Process.}, vol.~58, no.~8, pp. 4167--4181, Aug. 2010.

\bibitem{wang2016coarrays}
M.~Wang and A.~Nehorai, ``{Coarrays, MUSIC, and the Cram{\'e}r--Rao bound},''
  \emph{IEEE Trans. Signal Process.}, vol.~65, no.~4, pp. 933--946, Feb. 2017.

\bibitem{zeng2024tutorial}
Y.~Zeng, J.~Chen, J.~Xu, D.~Wu, X.~Xu, S.~Jin, X.~Gao, D.~Gesbert, S.~Cui, and
  R.~Zhang, ``A tutorial on environment-aware communications via channel
  knowledge map for {6G},'' \emph{IEEE Commun. Surv. Tuts.}, vol.~26, no.~3,
  pp. 1478--1519, 3rd Quart. 2024.

\bibitem{wu2021intelligent}
Q.~Wu, S.~Zhang, B.~Zheng, C.~You, and R.~Zhang, ``Intelligent reflecting
  surface-aided wireless communications: A tutorial,'' \emph{IEEE Trans.
  Commun.}, vol.~69, no.~5, pp. 3313--3351, May 2021.

\bibitem{di2020smart}
M.~Di~Renzo, A.~Zappone, M.~Debbah, M.-S. Alouini, C.~Yuen, J.~De~Rosny, and
  S.~Tretyakov, ``Smart radio environments empowered by reconfigurable
  intelligent surfaces: How it works, state of research, and the road ahead,''
  \emph{IEEE J. Sel. Areas Commun.}, vol.~38, no.~11, pp. 2450--2525, Nov.
  2020.

\bibitem{lu2021aerial}
H.~Lu, Y.~Zeng, S.~Jin, and R.~Zhang, ``Aerial intelligent reflecting surface:
  Joint placement and passive beamforming design with {3D} beam flattening,''
  \emph{IEEE Trans. Wireless Commun.}, vol.~20, no.~7, pp. 4128--4143, Jul.
  2021.

\bibitem{tang2020wireless}
W.~Tang, M.~Z. Chen, X.~Chen, J.~Y. Dai, Y.~Han, M.~Di~Renzo, Y.~Zeng, S.~Jin,
  Q.~Cheng, and T.~J. Cui, ``Wireless communications with reconfigurable
  intelligent surface: Path loss modeling and experimental measurement,''
  \emph{IEEE Trans. Wireless Commun.}, vol.~20, no.~1, pp. 421--439, 2020.

\bibitem{ryerson1960passive}
J.~Ryerson, ``Passive satellite communication,'' \emph{Proc. IRE}, vol.~48,
  no.~4, pp. 613--619, Apr. 1960.

\bibitem{cutler1965passive}
C.~C. Cutler, ``Passive repeaters for satellite communication systems,'' Feb.
  1965, {U.S. Patent 3 169 245}.

\bibitem{rahmat2015reflector}
Y.~Rahmat-Samii and R.~Haupt, ``Reflector antenna developments: a perspective
  on the past, present and future,'' \emph{IEEE Antennas Propag. Mag.},
  vol.~57, no.~2, pp. 85--95, Apr. 2015.

\bibitem{balanis2016antenna}
C.~A. Balanis, \emph{Antenna theory: Analysis and design}.\hskip 1em plus 0.5em
  minus 0.4em\relax Hoboken, NJ, USA: Wiley, 2016.

\bibitem{huang2004investigation}
Y.~Huang, N.~Yi, and X.~Zhu, ``Investigation of using passive repeaters for
  indoor radio coverage improvement,'' in \emph{Proc. IEEE Int. Symp. Antennas
  Propag.}, vol.~2, Jun. 2004, pp. 1623--1626.

\bibitem{chan20153d}
J.~Chan, C.~Zheng, and X.~Zhou, ``{3D printing your wireless coverage},'' in
  \emph{Proc. ACM Int. Workshop Hot Topics Wireless (HotWireless)}, Sep. 2015,
  pp. 1--5.

\bibitem{han2017enhancing}
S.~Han and K.~G. Shin, ``Enhancing wireless performance using reflectors,'' in
  \emph{Proc. IEEE INFOCOM-IEEE Conf. Comput. Commun.}, May 2017, pp. 1--9.

\bibitem{khawaja2020coverage}
W.~Khawaja, O.~Ozdemir, Y.~Yapici, F.~Erden, and I.~Guvenc, ``{Coverage
  enhancement for NLOS mmWave links using passive reflectors},'' \emph{IEEE
  Open J. Commun. Soc.}, vol.~1, pp. 263--281, Jan. 2020.

\bibitem{hager2023holistic}
S.~H{\"a}ger, K.~Heimann, S.~B{\"o}cker, and C.~Wietfeld, ``Holistic
  enlightening of blackspots with passive tailorable reflecting surfaces for
  efficient urban {mmWave} networks,'' \emph{IEEE Access}, vol.~11, pp.
  39\,318--39\,332, Apr. 2023.

\bibitem{anjinappa2021base}
C.~K. Anjinappa, F.~Erden, and I.~G{\"u}ven{\c{c}}, ``{Base station and passive
  reflectors placement for urban mmWave networks},'' \emph{IEEE Trans. Veh.
  Technol.}, vol.~70, no.~4, pp. 3525--3539, Apr. 2021.

\bibitem{singh2023stabilizing}
S.~Singh, H.~Tran, and T.~Le, ``{Stabilizing terahertz MIMO channel capacity
  with controlled diffuse reflections},'' in \emph{Proc. IEEE Int. Conf.
  Commun. (ICC)}, May 2023, pp. 5824--5830.

\bibitem{yu2023wireless}
Z.~Yu, C.~Feng, Y.~Zeng, T.~Li, and S.~Jin, ``Wireless communication using
  metal reflectors: Reflection modelling and experimental verification,'' in
  \emph{Proc. IEEE Int. Conf. Commun. (ICC)}, May 2023, pp. 4701--4706.

\bibitem{barreiro2006passive}
J.~L. D. L.~T. Barreiro and F.~L.~E. Azpiroz, ``Passive reflector for a mobile
  communication device,'' Aug. 2006, {U.S. Patent 7 084 819}.

\bibitem{ozdogan2020intelligent}
{\"O}.~{\"O}zdogan, E.~Bj{\"o}rnson, and E.~G. Larsson, ``Intelligent
  reflecting surfaces: Physics, propagation, and pathloss modeling,''
  \emph{IEEE Wireless Commun. Lett.}, vol.~9, no.~5, pp. 581--585, May 2020.

\bibitem{zhu2024movable}
L.~Zhu, W.~Ma, and R.~Zhang, ``Movable antennas for wireless communication:
  Opportunities and challenges,'' \emph{IEEE Commun. Mag.}, vol.~62, no.~6, pp.
  114--120, Jun. 2024.

\bibitem{zhu2024modeling}
------, ``Modeling and performance analysis for movable antenna enabled
  wireless communications,'' \emph{IEEE Trans. Wireless Commun.}, vol.~23,
  no.~6, pp. 6234--6250, Jun. 2024.

\bibitem{ma2024mimo}
W.~Ma, L.~Zhu, and R.~Zhang, ``{MIMO capacity characterization for movable
  antenna systems},'' \emph{IEEE Trans. Wireless Commun.}, vol.~23, no.~4, pp.
  3392--3407, Apr. 2024.

\bibitem{zhu2025tutorial}
L.~Zhu \emph{et~al.}, ``A tutorial on movable antennas for wireless networks,''
  \emph{IEEE Commun. Surv. Tuts.}, 2025, doi: 10.1109/COMST.2025.3546373.

\bibitem{shao20246d}
X.~Shao, Q.~Jiang, and R.~Zhang, ``{6D} movable antenna based on user
  distribution: Modeling and optimization,'' \emph{IEEE Trans. Wireless
  Commun.}, vol.~24, no.~1, pp. 355--370, Jan. 2025.

\bibitem{shao20246ddiscrete}
X.~Shao, R.~Zhang, Q.~Jiang, and R.~Schober, ``{6D} movable antenna enhanced
  wireless network via discrete position and rotation optimization,''
  \emph{IEEE J. Sel. Areas Commun.}, vol.~43, no.~3, pp. 674--687, Mar. 2025.

\bibitem{shao20246dma}
X.~Shao and R.~Zhang, ``{6DMA} enhanced wireless network with flexible antenna
  position and rotation: Opportunities and challenges,'' \emph{IEEE Commun.
  Mag.}, 2024.

\bibitem{shao2025distributed}
X.~Shao, R.~Zhang, Q.~Jiang, J.~Park, T.~Q. Quek, and R.~Schober, ``Distributed
  channel estimation and optimization for {6D} movable antenna: Unveiling
  directional sparsity,'' \emph{IEEE J. Sel. Topics Signal Process.}, 2025,
  doi: 10.1109/JSTSP.2025.3539085.

\bibitem{ning2024movable}
B.~Ning, S.~Yang, Y.~Wu, P.~Wang, W.~Mei, C.~Yuen, and E.~Bj{\"o}rnson,
  ``Movable antenna-enhanced wireless communications: General architectures and
  implementation methods,'' \emph{arXiv preprint arXiv:2407.15448}, 2024.

\bibitem{zheng2025rotatable}
B.~Zheng, Q.~Wu, and R.~Zhang, ``Rotatable antenna enabled wireless
  communication: Modeling and optimization,'' \emph{arXiv preprint
  arXiv:2501.02595}, 2025.

\bibitem{diebel2006representing}
J.~Diebel \emph{et~al.}, ``Representing attitude: Euler angles, unit
  quaternions, and rotation vectors,'' \emph{Matrix}, vol.~58, no. 15-16, pp.
  1--35, Oct. 2006.

\end{thebibliography}

\end{document}